\documentclass[sigconf]{acmart}

\AtBeginDocument{%
  \providecommand\BibTeX{{%
    \normalfont B\kern-0.5em{\scshape i\kern-0.25em b}\kern-0.8em\TeX}}}

\setcopyright{acmcopyright}
\copyrightyear{2024}
\acmYear{2024}
\setcopyright{acmlicensed}\acmConference[CHI '24]{Proceedings of the CHI Conference on Human Factors in Computing Systems}{May 11--16, 2024}{Honolulu, HI, USA}
\acmBooktitle{Proceedings of the CHI Conference on Human Factors in Computing Systems (CHI '24), May 11--16, 2024, Honolulu, HI, USA}
\acmDOI{10.1145/3613904.3642152}
\acmISBN{979-8-4007-0330-0/24/05}

\acmSubmissionID{8018}

\usepackage{multirow}
\usepackage{color}
\usepackage{colortbl}
\usepackage{stfloats}
\usepackage{longfbox} 
\usepackage{subcaption}
\usepackage{enumitem}
\usepackage{adjustbox}
\usepackage{graphicx}
\usepackage{subcaption}
\usepackage{tabularx}

\newcommand{\sysname}{\textsc{ChaCha}}

\makeatletter
\newdimen\@tempdimd
\makeatother

\newcommand{\eg}{\textit{e.g.}}
\newcommand{\ie}{\textit{i.e.}}
\newcommand{\cf}{\textit{c.f.}}
\newcommand{\etal}{\textit{et al.}}

\newfboxstyle{patternparam}{padding=2pt, padding-left=4pt, padding-right=4pt, border-style=none, height=6pt, border-radius=5pt}

\newcommand{\rectwrap}[2]{\lfbox[patternparam, border-radius=0pt, padding-left=2.5pt, padding-right=2.5pt, background-color=#1]{\sffamily{\textcolor{white}{#2}}}}

\newcommand{\rectwrapsmall}[2]{\lfbox[patternparam, border-radius=0pt, padding-left=2pt, padding-right=2pt, height=5pt, background-color=#1]{\sffamily{\textcolor{white}{#2}}}}

\newcommand{\phaseexplore}{\lfbox[patternparam, background-color=explorecolor]{\sffamily{\textcolor{white}{Explore}}}}

\newcommand{\phaselabel}{\lfbox[patternparam, background-color=labelcolor]{\sffamily{\textcolor{white}{Label}}}}

\newcommand{\phasefind}{\lfbox[patternparam, background-color=findcolor]{\sffamily{\textcolor{white}{Find}}}}

\newcommand{\phaserecord}{\lfbox[patternparam, background-color=recordcolor]{\sffamily{\textcolor{white}{Record}}}}

\newcommand{\phaseshare}{\lfbox[patternparam, background-color=sharecolor]{\sffamily{\textcolor{white}{Share}}}}

\newcommand{\phaseexplorelg}{\lfbox[patternparam, height=9pt, border-radius=6pt, background-color=explorecolor]{\sffamily{\textcolor{white}{Explore}}}}

\newcommand{\phaselabellg}{\lfbox[patternparam, height=9pt, border-radius=6pt, background-color=labelcolor]{\sffamily{\textcolor{white}{Label}}}}

\newcommand{\phasefindlg}{\lfbox[patternparam, height=9pt, border-radius=6pt, background-color=findcolor]{\sffamily{\textcolor{white}{Find}}}}

\newcommand{\phaserecordlg}{\lfbox[patternparam, height=9pt, border-radius=6pt, background-color=recordcolor]{\sffamily{\textcolor{white}{Record}}}}

\newcommand{\phasesharelg}{\lfbox[patternparam, height=9pt, border-radius=6pt, background-color=sharecolor]{\sffamily{\textcolor{white}{Share}}}}

\newcommand{\testexplore}{\rectwrap{explorecolor}{T1}}

\newcommand{\testlabel}{\rectwrap{labelcolor}{T2}}

\newcommand{\testrecord}{\rectwrap{recordcolor}{T4}}

\newcommand{\blackrectsmall}[1]{\rectwrapsmall{darkgray}{#1}}

\definecolor{explorecolor}{HTML}{9D62A8}
\definecolor{labelcolor}{HTML}{90C73E}
\definecolor{findcolor}{HTML}{CA5576}
\definecolor{recordcolor}{HTML}{A5684F}
\definecolor{sharecolor}{HTML}{5E95CF}

\definecolor{revisedcolor}{RGB}{0,0,255}

\newcommand{\revised}[1]{#1}

\newcommand{\ipstart}[1]{\vspace{1mm} \noindent{\textbf{\textit{#1.}}}}

\newcommand{\npstart}[1]{\vspace{1mm} \noindent{{#1}}}

\newcommand{\circledigit}[1]{\textbf{\normalsize{\textsf{\textcircled{\footnotesize{#1}}}}}}

\definecolor{tableheader}{HTML}{EFEFEF}
\definecolor{tablegrayline}{HTML}{d0d0d0}

\newcommand{\labelphantom}[1]{%
  \parbox{0pt}{\phantomsubcaption\label{#1}}%
}

\newenvironment{quotetable}{
\vspace{3mm}
\hfill\break
\noindent
    \small\sffamily\centering
    \def\arraystretch{1.15}\setlength{\tabcolsep}{0.25em}
    \tabularx{\columnwidth}{c>{\raggedright\let\newline\\\arraybackslash\hspace{0pt}}m{0.91\columnwidth}}
    \hline}
{
    \arrayrulecolor{black}\hline
    \endtabularx
    \vspace{3mm}
}

\newcommand{\quotebotline}[1]{\textbf{\raisebox{-4pt}{\includegraphics[width=12pt]{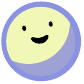}}} & \textit{#1}\\\arrayrulecolor{tablegrayline}\hline}

\newcommand{\quoteuserline}[2]{\textbf{#1} & \textit{#2}\\\arrayrulecolor{tablegrayline}\hline}

\newcommand{\symboldescribe}{\raisebox{-2pt}{\includegraphics[width=8.5pt]{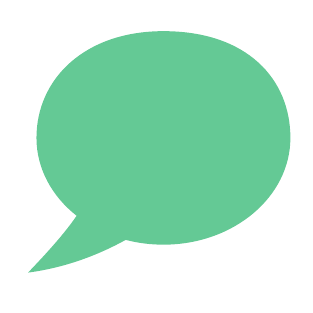}} }
\newcommand{\symbolselect}{\raisebox{-2pt}{\includegraphics[width=8pt]{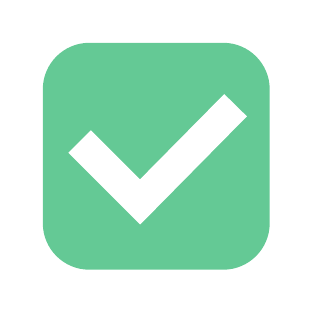}} }

\newfboxstyle{emotiontagparam}{padding=2pt, padding-left=3pt, padding-right=3pt, height=7pt, border-radius=2pt, margin-top=1pt, border-color=lightgray}

\newcommand{\emotiondescribed}[1]{\lfbox[emotiontagparam]{\symboldescribe~#1}}
\newcommand{\emotionselected}[1]{\lfbox[emotiontagparam]{\symbolselect~#1}}

\newcommand{\emojismile}{\raisebox{-1pt}{\includegraphics[width=9pt]{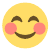}}}

\begin{document}

\title{\sysname: Leveraging Large Language Models to Prompt Children~to Share Their Emotions about Personal Events}

\author{Woosuk Seo}
\authornote{Woosuk Seo conducted this work as a research intern at NAVER AI Lab.}
\orcid{0000-0002-1316-4341}
\affiliation{%
  \institution{University of Michigan}
  \city{Ann Arbor}
  \country{USA}
}
\email{seow@umich.edu}

\author{Chanmo Yang}
\orcid{0000-0002-4959-7595}
\affiliation{%
  \institution{Wonkwang Univ. Hospital \& Wonkwang University}
  \country{Republic of Korea}
}
\email{ychanmo@wku.ac.kr}

\author{Young-Ho Kim}
\orcid{0000-0002-2681-2774}
\affiliation{%
  \institution{NAVER AI Lab}
  \country{Republic of Korea}
}
\email{yghokim@younghokim.net}

\renewcommand{\shortauthors}{Woosuk Seo, Chanmo Yang, and Young-Ho Kim}

\begin{abstract}
Children typically learn to identify and express their emotions by sharing stories and feelings with others, particularly family members. However, it is challenging for parents or siblings to have effective emotion communication with children since children are still developing their communication skills. We present \sysname{}, a chatbot that encourages and guides children to share personal events and associated emotions. \sysname{} combines a state machine and large language models (LLMs) to keep the dialogue on track while carrying on free-form conversations. Through an exploratory study with 20 children (aged 8--12), we examine how \sysname{} prompts children to share personal events and guides them to describe associated emotions. Participants perceived \sysname{} as a close friend and shared their stories on various topics, such as family trips and personal achievements. Based on the findings, we discuss opportunities for leveraging LLMs to design child-friendly chatbots to support children in sharing emotions.
\end{abstract}



\begin{teaserfigure}
    \centering
    \includegraphics[width=\textwidth]{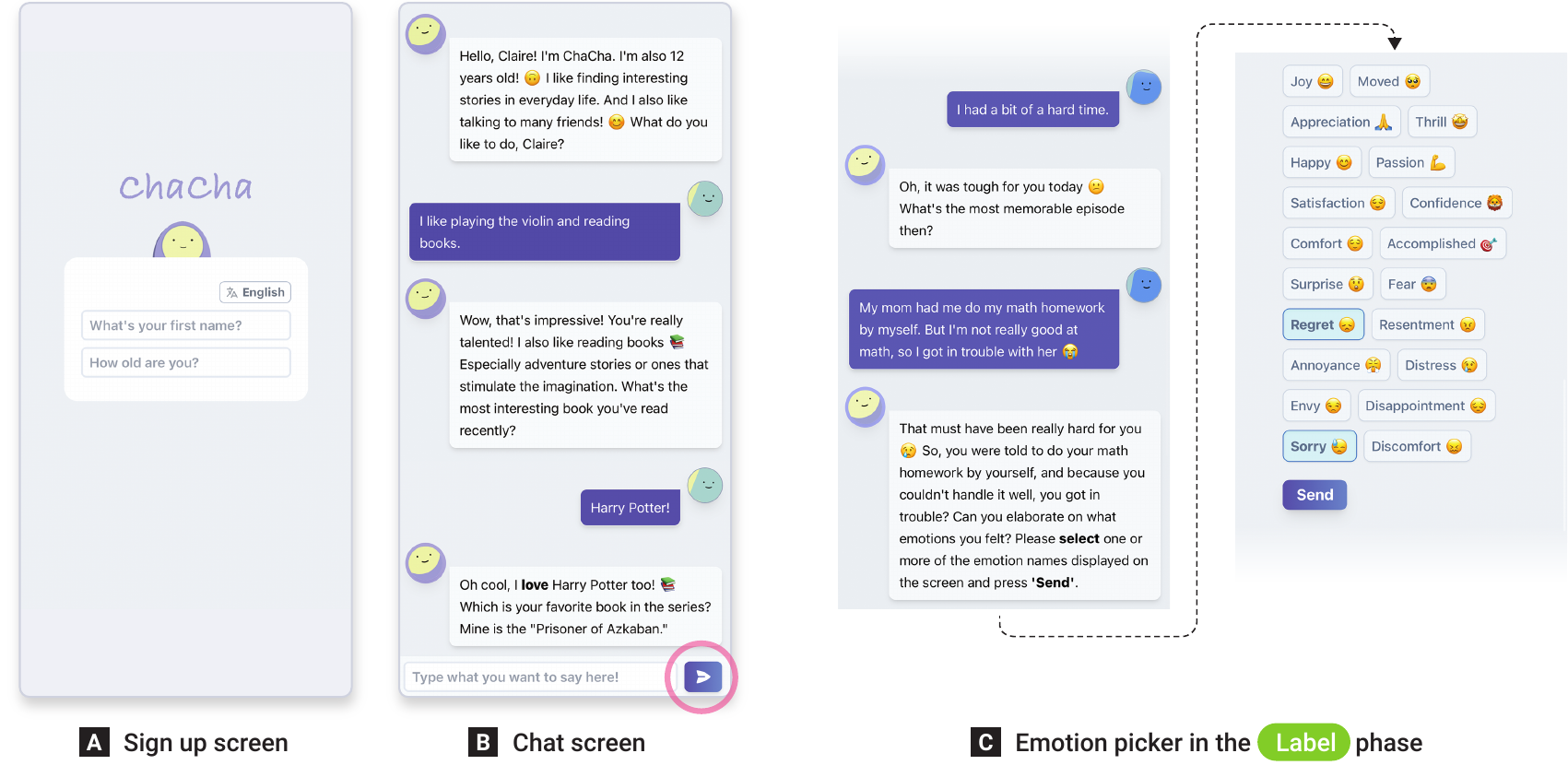}
    \vspace{-5mm}
    \Description[Three screen captures of ChaCha]{Three screen captures show the main screens of the ChaCha interface. The first scene shows how users can initiate a new session by entering their name and age. The second screen shows the user's conversation with ChaCha. The last screen shows how ChaCha provides them with a list of emotions for guidance when the user struggles with describing their emotions.}
    \caption{Main screens of \sysname{} interface. \blackrectsmall{A} The user can initiate a new session by entering their name and age, and \blackrectsmall{B} converse with \sysname{}. \blackrectsmall{C} If the user struggles with describing their emotions, \sysname{} offers a list of emotions for guidance.}
    \label{fig:interface:teaser}
\end{teaserfigure}

\maketitle

\section{Introduction}
In the context of child development, the perception and expression of emotions during early childhood hold significant importance. As children progress through different developmental stages, they become better at self-reporting their emotional experiences; they gradually develop the ability to express their emotions or withhold emotional expression to avoid potentially adverse reactions from others~\cite{saarni2007emotional,zeman1996display}. Hence, children need developmentally appropriate education and practice to develop such emotional competencies. Children's relationships with their parents or siblings significantly impact their emotional development in the family context. Better quality of sibling relationship is more likely to promote children's higher social-emotional well-being~\cite{yucel2015siblings}. Similarly, parents play an essential role in supporting how children identify and express their emotions ~\cite{brumariu2015mother, hernandez2021toward}. For instance, mothers' behaviors in providing emotional support or controlling their children's emotional states significantly impacted how children manage their anxiety ~\cite{brumariu2015mother}.


Despite its significance, emotion communication \footnote{Hereinafter, we define emotion communication as the process of recognizing and sharing emotions between two or more individuals.} has not been frequently addressed in parenting interventions~\cite{shaffer2019let}. For parents, it is challenging to guide emotion communication with children who are still developing communication skills ~\cite{shaffer2019let}. With the lack of support and interventions, parents would provide less emotional support, which may result in adverse mental health outcomes for children, such as anxiety ~\cite{brumariu2015mother}. Notably, only-child children may have more challenges than those with siblings since they would have fewer opportunities to build supportive sibling relationships and promote their emotional well-being (\eg, ~\cite{yucel2015siblings}). Even worse, the COVID-19 pandemic has impacted children's social activities and daily routines, affecting their socio-emotional development ~\cite{egan2021missing,hen2022children}: A recent survey study showed that, along with anxiety and emotional problems, children and adolescents (ages 8--17) have experienced more difficulty in emotional regulation during the pandemic~\cite{hen2022children}. 

In the HCI community, prior studies explored how technology can be designed to support children with learning and sharing their emotions ~\cite{ryu2021dot,santos2020therapist}
or to enhance parents' awareness of children's emotional well-being ~\cite{pepping2020motiis}. These studies showed how a chatbot~\cite{santos2020therapist} or a tangible artifact~\cite{pepping2020motiis} can be used to detect children's emotions more accurately. For instance, Santos \etal{} ~\cite{santos2020therapist} demonstrated how a chatbot leverages a storytelling strategy to effectively recognize children's emotions from the given list of emotions. Although prior studies have presented opportunities for technology to promote children's emotional regulation through detection, they primarily focused on collecting data from children and sharing it with their parents, limiting our understanding of how technology can better support children in identifying and expressing their emotions. The existing technology was mostly limited to probing questions to recognize children's emotions rather than supporting them with practicing to express their emotions. Children’s perceptions and preferences for how, when, and what to communicate about their emotions were often overlooked. 

To explore how chatbots could support children in identifying and expressing their emotions, we present \sysname{} (\autoref{fig:interface:teaser}), a chatbot that guides children in sharing their emotions about personal events. 
Following the emotion coaching guidelines~\cite{gottman1998raising}, \sysname{} elicits a child's key events, helps them label associated emotions, explores possible solutions to alleviate adverse emotional experiences, and encourages sharing their emotions with parents.
For better engagement with children in the conversation, \sysname{}'s dialogue system leverages large language models (LLMs)~\cite{Brown2020FewShotLearners, openai2023gpt4} to coherently follow up the conversation context while generating \textit{in-situ} responses on serendipitous topics. Informed by formative interviews ($N=6$) with child mental health professionals (\eg{}, psychiatrists, psychotherapists), we designed the conversations to consist of multiple phases with sub-goals and a dedicated LLM to keep the LLM-driven conversation on track and easy for children to follow. With this LLM-driven chatbot, we aim to answer the following research question: \textit{How feasible is an LLM-driven chatbot in prompting children to share their emotions about personal events?}

To examine how children interact with \sysname{} and perceive it for sharing their emotions about personal events, we conducted an exploratory user study with 20 children in South Korea. During the study session, participants freely conversed with \sysname{} on a smartphone for about 30 minutes. We observed that participants felt comfortable conversing with \sysname{}, sharing diverse events ranging from a family trip to a conflict with their parents. With \sysname{}'s assistance, participants were able to recall and identify their emotions associated with the positive and negative events. Participants also recognized \sysname{} was skilled in listening to their stories about events and empathizing with their emotions, hence perceived \sysname{} as a close friend with whom they would share emotions that they would not share with others.

\npstart{In summary, this work offers the following contributions:}
\begin{enumerate}[leftmargin=*, itemsep=4pt, topsep=0pt]
\item The design and implementation of \sysname{}, a novel LLM-driven chatbot that helps children share their emotions about events through free-form conversations. \sysname{}'s design was informed by formative interviews ($N=6$) with children's mental health professionals. We present design rationales of a chatbot for encouraging children to express their emotions and conversational flow of \sysname{} to meet those rationales. \revised{The source code of \sysname{} is available at \urlstyle{tt}\url{https://naver-ai.github.io/chacha}.}
\item An empirical understanding of how children recognize and interact with \sysname{}. We conducted an exploratory study with 20 children. From \revised{the analysis of the dialogue logs and interview transcripts}, we provide findings on how children interact with a chatbot driven by LLMs in sharing emotions about personal events.
\item Design considerations for LLM-driven chatbots for children. We present the benefits of leveraging LLMs in designing child-friendly chatbots, such as showing empathetic behaviors that encourage children to share their emotions, and suggestions to mitigate potential concerns about using LLMs, such as balancing chatbots' self-disclosure and overreliance on chatbots.
\end{enumerate}


\section{Related Work}
In this section, we explore prior studies about supporting children's emotional development and communication and examine existing technology for children's emotion regulation and mental health. We then review previous work about the characteristics of LLM-driven chatbots, presenting the potential advantages of LLMs in children's mental health. 

\subsection{Emotion Communication and Coaching}
The abilities to identify and express emotions are two essential skills for emotional competence discussed in the field of children's emotional development~\cite{saarni2007emotional}. From infants, children begin to learn expressive behaviors that are emotion-related components~\cite{Saarni_2022}. Along with cognitive and social skills, emotional competence can be developed with proper support and age-appropriate education as children gradually acquire more knowledge and experiences. Due to this nature of emotional development, parents' emotion communication with children essentially affects the development of children's skills to recognize, express, and share their emotions.
Ample literature in psychology and medicine has emphasized how children's emotional development can be affected by how parents communicate about children's emotions~\cite{brumariu2015mother, hernandez2021toward,denham1995mother,ashiabi2000promoting}. In particular, middle childhood (ages 6--12) is considered a necessary time to foster open communication with parents. Children in this age range are old enough to decide how to express which emotions to with others \cite{saarni1999development}, while they still need to practice and develop skills to express their opinions accurately \cite{piaget1952origins}. In addition, middle childhood is when children spend more time away from their parents than younger children. As a result, parents would have less awareness of their children's negative experiences unless children choose to discuss them~\cite{gentzler2005parent}. \revised{Parents may face considerable challenges in providing adequate support to their children when they lack an understanding of children's feelings~\cite{seo2021challenges}.}

To support parents in facilitating emotion communication and regulation of children, a renowned psychologist, John Gottman, developed a guideline for emotion coaching to help parents teach their children how to identify and express their emotions~\cite{gottman1998raising}. The guideline consists of five steps: (1)~Be aware of the child's emotions, (2)~Recognize the child's expression of emotions, (3)~Listen with empathy, (4)~Help the child learn to label their emotions, and 
 (5)~Set limits when helping the child to solve problems~\cite{gottman1998raising}. Although emotion coaching is a widely used and validated framework~\cite{gus2015emotion}, parents often face various challenges in thoroughly learning and applying the principles of emotion coaching to their contexts~\cite{Havighurst2021TuningInToKids}. Such challenges may include a lack of experience with emotion communication in their own childhood, a lack of verbal skills to name the emotions that their children express, and fear of letting their children verbalize specific emotion  words~\cite{Havighurst2021TuningInToKids}. To complement parents' efforts in emotion communication with children, prior studies have proposed interventions for children to support them in sharing their experiences and emotions~\cite{havighurst2015emotion, Havighurst2021TuningInToKids}.

\subsection{Chatbots for Children's Emotion Regulation and Mental Health}
Technological interventions have supported children with emotion regulation and mental health (\eg, mHealth apps~\cite{nicolaidou2022gamified, moltrecht2022transdiagnostic}). Among various types of technology, chatbots have lately shown promising opportunities in promoting emotion regulation and mental health of children and adolescents ~\cite{tsai2019employing, santos2020therapist, dosovitsky2023development}. Through conversational interactions, chatbots offer a space where children can freely share their stories and express their emotions in their own words while keeping their secrets from others \cite{kim2018can}. 


Acknowledging the benefits of conversational interactions, prior work has proposed chatbots as a promising way to assist children's emotion regulation or promote their mental health. Dosovitsky~\etal{} \cite{dosovitsky2023development} developed BethBot, a chatbot that includes cognitive behavior therapy (CBT) intervention modules used in psychoeducation to educate adolescents on depression and teach coping skills. Based on the analysis of user experience questions, Dosovitsky~\etal{} \revised{\cite{dosovitsky2023development}} found that adolescents perceived BethBot as an acceptable mental health resource that would improve their symptoms. Santos~\etal{} \revised{\cite{santos2020therapist}} developed a chatbot by adapting a storytelling strategy. The chatbot recognizes specific keywords from children's responses to determine their emotions on the given list of emotions. The user testing with children (ages 9-11) revealed that the child participants felt comfortable sharing their stories with the chatbot and showed how accurately the chatbot could detect children's emotions based on their storytelling. Although prior work has posed the benefits of leveraging chatbots for children, existing chatbot research in this domain predominantly adopts a rule-based approach~\cite{abd2019overview} where the chatbot follows a predefined conversational flow with close-ended questions and templated messages. Hence, the chatbot may not be responsive to children's messages that are beyond the scope of the designer's prior considerations~\cite{Huang2020ChallengesODD, abd2019overview}. Given the open-ended nature of sharing emotions and personal events, we believe that free-form conversations could be more effective for carrying on communication about children's emotions. Thus, our study extends the line of research about chatbots for children's mental health by demonstrating the opportunity for LLMs to enhance the open-ended conversation capability of a chatbot.

\subsection{Large Language Models for Conversation}
The recent advance of pre-trained LLMs (\eg{}, GPT~\cite{Brown2020FewShotLearners, openai2023gpt4}, PaLM~\cite{Chowdhery2022PALM}, LLaMA~\cite{touvron2023llama}, LaMDA~\cite{thoppilan2022lamda},  HyperCLOVA~\cite{Kim2021HyperCLOVA}) has presented new opportunities for designing and improving various natural language processing tasks~\cite{zhao2023surveyLLM, Lie2023PreTrainPrompt}. 
Specifically, LLMs led to the invention of a new design approach for chatbots, complementing the limitations of traditional rule-based and retrieval-based approaches~\cite{Huang2020ChallengesODD, wei2023leveraging, jo2023carecallinterview}. When used to bootstrap a chatbot, LLMs generate a message that naturally continues the current dialogue by augmenting the content generation process with their large-scale pre-trained knowledge. Hence, the LLM-generated messages tend to be coherent across the conversation session and human-like, which is more flexible than rule-based chatbots. Unlike retrieval-based chatbots that require a number of domain-specific dialogue data to train a retriever~\cite{Huang2020ChallengesODD}, only a few examples or even no examples can bootstrap comparably working chatbots~\cite{wei2023leveraging}. For instance, Wei~\etal{} presented GPT-3-driven chatbots devised with simple zero-shot prompts to collect health-related questionnaires through free-form conversation~\cite{wei2023leveraging}. More recently, enterprises introduced LLM-driven chatbot services (\eg, ChatGPT~\cite{chatgpt}, Bard~\cite{Bard}, HyperClovaX~\cite{clovax2023}, Pi~\cite{PiChatbot}) that enable individuals to perform a variety of tasks, such as idea generation and data processing, via conversational interactions~\cite{dowling2023chatgpt}.

Adopting this benefit of leveraging LLMs, recent studies have explored the potential use of LLMs in mental health contexts (\eg, ChatGPT-generated patient-psychiatrist simulations ~\cite{chen2023llm}). LLMs may also express personality traits ~\cite{jiang2023personallm}, maximizing their content generation capability for a specific context and target. This aspect of LLMs could be helpful in facilitating a free-form conversation about emotional distress or psychological issues by simulating a more sociable, attentive, or caring personality. However, further studies are required to examine the feasibility of this relatively new technology in promoting mental health, particularly for children. A recent study presented how an LLM-driven agent can help children generate curious questions, potentially suitable for training their question-asking skills ~\cite{abdelghani2022gpt}. Similar to this case of using LLM for skill development, LLM-driven conversations could provide a space for children to communicate about their emotions more openly rather than simply answering questions by rule-based chatbots. Such open communication in free-form conversation could help children develop their skills to identify and express their emotions in their own words. Motivated by such potential of LLMs, we examine how an LLM-driven chatbot can prompt children to share their emotions about specific events they experience. 

\section{Formative Interviews}

\begin{figure*}[b]
    \centering
     \begin{subfigure}[b]{0.45\textwidth}
         \centering
         \includegraphics[height=5cm]{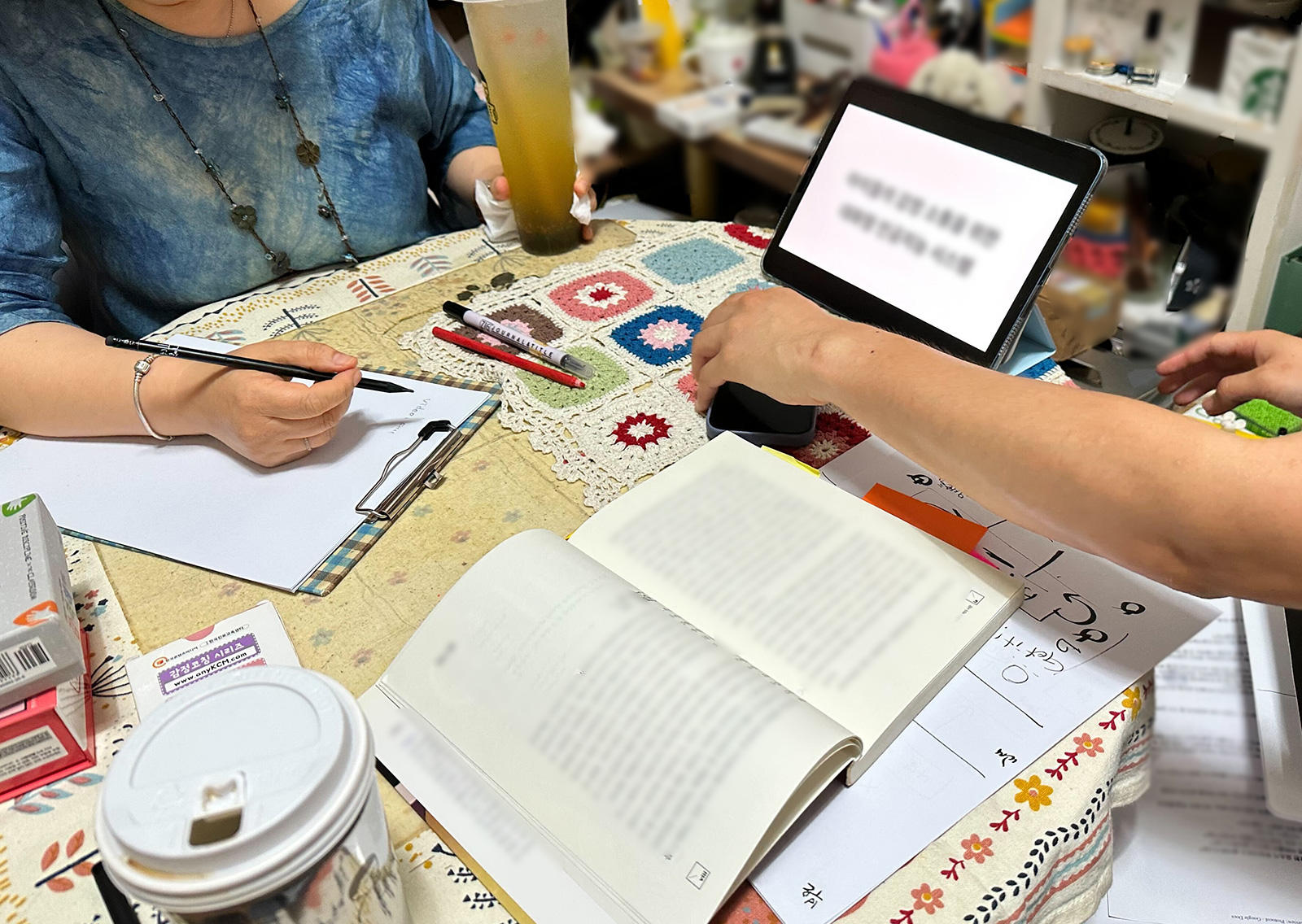}
         \caption{Interview settings}         \label{fig:formative:photos:settings}
     \end{subfigure}
     \hspace{3mm}
     \begin{subfigure}[b]{0.45\textwidth}
         \centering
         \includegraphics[height=5cm]{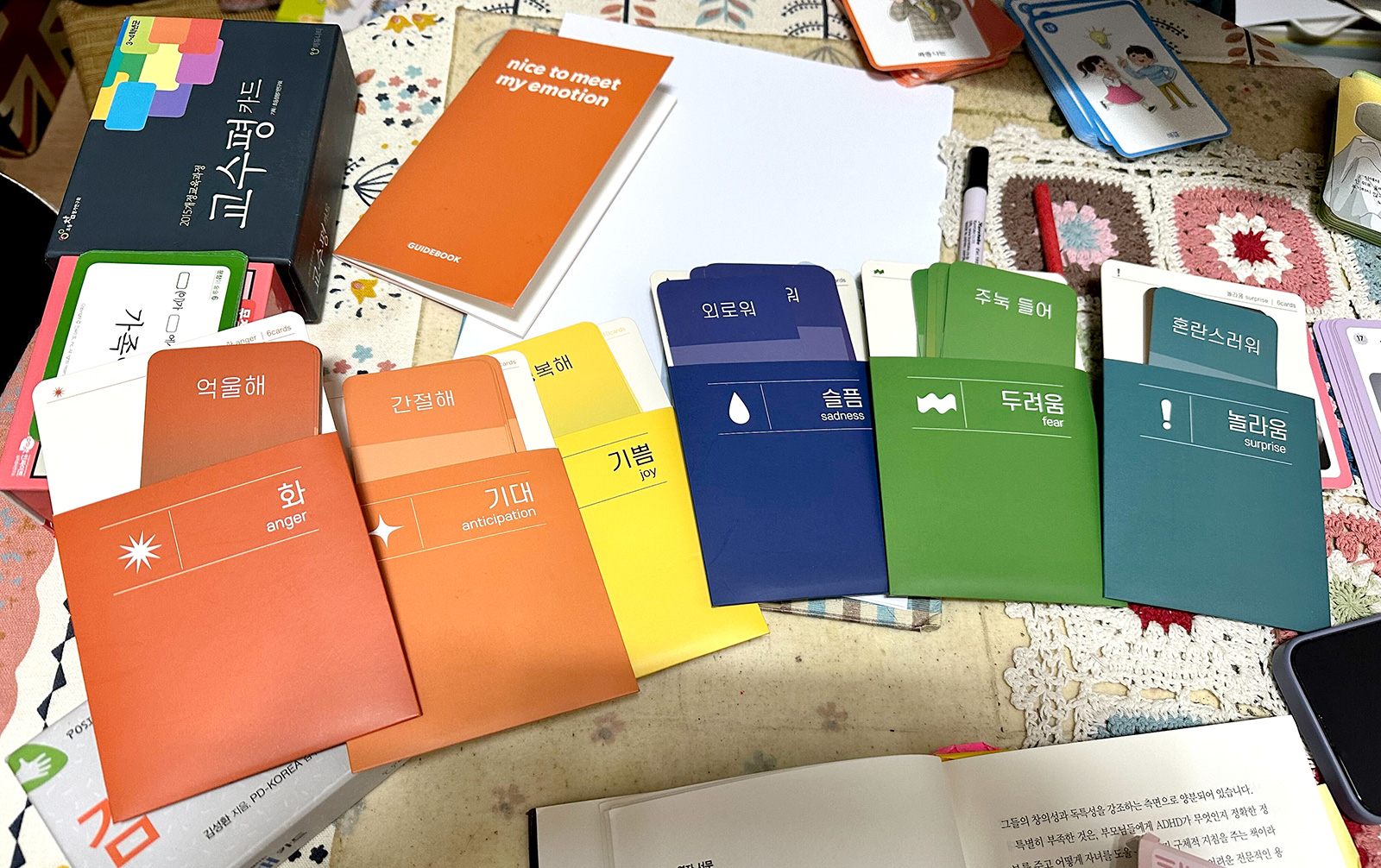}
         \caption{Example emotion cards}
         \label{fig:formative:photos:cards}
     \end{subfigure}
    \Description[Two photos taken in one interviewee’s office for formative study]{Two photos taken in one interviewee’s office for formative study. The first photo shows the interview settings with a book that the interviewee showed and a tablet for demonstrating a video prototype. The second photo shows emotion cards that the interviewee uses when guiding children to describe their emotions. The emotion keywords on the envelopes are anger, anticipation, joy, sadness, fear, and surprise (from left to right)}
    \caption{Photos taken in one interviewee's office: (a) The interview settings with a book that the interviewee showed and a tablet of the experimenter to demonstrate a video prototype, and (b) emotion cards that the interviewee uses when guiding children to describe their emotions. The emotion keywords on the envelopes are anger, anticipation, joy, sadness, fear, and surprise (from left to right).}
    \label{fig:formative:photos}
\end{figure*}

To understand clinical practices for communicating emotions with children, we conducted semi-structured interviews with six child mental health professionals based in South Korea. The interviewees were three child psychiatrists, two child psychotherapists, and one child development specialist. We recruited them from university hospitals, child psychotherapy centers, and a private child mental health clinic. The interviews lasted about 1 to 1.5 hours, and two researchers joined the interviews. The interviews were conducted in person or remotely based on each professional's availability. We offered 100,000 KRW (approx. 80 USD) as compensation.  

We asked about the professionals' perspectives on the challenges children face when sharing their emotions, strategies to better communicate with them about their feelings, and a chatbot that supports children in sharing emotions. As a probe, we showed the professionals a video prototype of a conversation between a chatbot and a child (\autoref{fig:formative:photos:settings}; See supplementary material). The purpose of the video was to help interviewees understand the expected conversations between children and a chatbot. We also asked the interviewees to provide any materials or tools they refer to or use when communicating with children (See \autoref{fig:formative:photos:cards}). The interviews were audio recorded and transcribed. \revised{The first author open-coded transcripts, going through a few iterations. The full research team then iterated multiple sessions of discussions identifying themes regarding design considerations. Through this process, we finalized three design considerations based on professionals' approach to emotion communication with children.} 

We identified three considerations for designing a chatbot for emotion communication with children: First, \textbf{a chatbot should show its empathy with children's emotions.} The professionals articulated the importance of empathy in helping children define and express their feelings. A common strategy to empathize with children's feelings is to \textit{repeat what they have shared} (\eg, ``\textit{I see. You were sad when you lost the game}''). Specifically, it is important to empathize with children from their level of understanding, focusing on what children actually feel rather than judging their feelings. \revised{All of the professionals mentioned that it seems to be appropriate to have a peer persona for a chatbot. One of them particularly noted that a peer chatbot would allow children to open themselves and share their emotions more easily. Also, children would be more accepting of the suggestions that the peer chatbot provides to them.} Second, \textbf{a chatbot should give children options for emotions to choose from.} Some children, particularly younger ones, answer questions with just one word (\eg, yes/no, good/bad). \revised{To probe more questions in such cases, some professionals shared that they often give potential options. Adopting this approach,} chatbots could help children who express difficulty in describing their emotions by providing some options for emotional words that they can choose from. \revised{Some professionals also noted that these options for emotional words can be helpful to children to explore various emotions.} Lastly, \textbf{a chatbot should encourage children to share their feelings with their parents.} The professionals envisioned the ideal role of the chatbot should be a bridge or a facilitator between children and their parents, not replacing the parents' role in children's emotional development. Hence, the chatbot should not simply be the one who solves and teaches everything that children ask. It should guide the children to share their emotional needs with their parents or healthcare providers so that they can receive relevant support.

Along with the three design considerations described above, the findings from the interviews with the professionals informed us of the conversation flow of the chatbot and the initial ideas of the chatbot's functions, which we will describe in \autoref{sec:system}. To carefully design the conversational flows and the chatbot's behaviors, we invited one of the psychiatrists we interviewed as an author of this work and collaborated with him on the design of the chatbot and the study protocol.
\section{\sysname}\label{sec:system}
Informed by the formative interviews, we designed and developed \sysname{}, a chatbot to prompt children to share their stories about specific events and associated emotions. We improved the interface and the underlying dialogue system by running pilot sessions with two children recruited by word of mouth. This section describes the design rationales for \sysname{} and the dialogue system, along with implementation details.

\subsection{Conversation Design Rationales}
\ipstart{DR1. \revised{Empathize children from their perspectives}}\phantomsection{}\label{sec:dr1}
Our target audience is child users who experience the world differently from adults rather than lack competencies~\cite{bruckman2007hci}. 
\revised{
Especially, children (age 6-12) begin to develop competencies and preferences in sharing emotions ~\cite{saarni1999development}. As one of such
preferences, children prefer to share emotions, which tend to change from adults such as parents to their peers ~\cite{saarni1988children}. For instance, in a sample of children (age 6-14), older children would share emotions with their peers rather than with their parents, whereas younger children preferred adults to peers ~\cite{saarni1988children}. Aligning with the findings from prior work, the insights from our interviews with professionals also advocate for the use of a peer persona for the chatbot. 
}
Thus, we decided to provide children with a peer interaction experience. By setting \sysname{} as a peer child, we expected the users to have more comfortable conversations about their stories and emotions. Moreover, based on this child persona, we tried to control the range of information that \sysname{} shares with children. For instance, we gave specific instructions to \sysname{} to only answer the questions that are somewhat related to children's interests or their stories and direct them to ask their parents if they ask inappropriate questions (\cf, \autoref{sec:appendix:speaking}). We leveraged an LLM to determine the boundary of child-appropriateness. For safety considerations, we conducted internal tests on \sysname{}'s responses to questions necessitating parental guidance (e.g., regarding violence) prior to initiating the user study.

\ipstart{DR2. \revised{Support children explore a range of emotions}}\phantomsection{}\label{sec:dr2}
From the formative interviews, the professionals articulated the importance of \revised{allowing children to explore a range of different emotions}. More importantly, negative emotions may need to be probed further since those emotions usually involve various stakeholders and factors (\eg, other friends, recurring occasions). It is essential to explore negative emotional experiences more carefully than positive ones so that children can learn how to manage negative emotions or ask for help. \revised{Through sharing the event that caused the negative emotions, children may develop different strategies to change their behaviors and goals that ultimately help them manage and alleviate their negative emotions. ~\cite{davis2010metacognitive}.} Thus, we branch out two phases to handle each category of emotions: \phasefind{} for negative emotions and \phaserecord{} for positive emotions. More detailed descriptions of those phases are explained in \autoref{sec:design:conversation}. 

\ipstart{DR3. \revised{Guide children rather than judge}}\phantomsection{}\label{sec:dr3}
The professionals also highlighted that a common mistake parents make is their tendency to judge their children's behaviors and emotions. This is because parents often want to understand the context of what happened so that they can provide appropriate support or solve the issues for their children. Reflecting parental needs, some existing technology for children's emotion regulation focuses on detecting emotions by confirming with children about their emotions (\eg, ``\textit{You look sad. Is that right?}''~\cite{santos2020therapist}). In contrast, we aimed to elicit children's emotions by acknowledging them rather than judging them. Our focus was on empathizing with emotions associated with a specific event and guiding them to develop strategies for managing their emotions by themselves first, then requesting support from their parents.
Therefore, we gave specific instructions to \sysname{}, focusing on acknowledging child users before intervening (\eg, ``\texttt{Empathize the user's emotion by restating how they felt and share your own experience that is similar to the user's.}''). \revised{Once \sysname{} acknowledges children's emotions, \sysname{} would guide child users through a structured conversation flow, which will be explained in detail in the following sub-section.}


\subsection{Conversation Design}\label{sec:design:conversation}

\begin{figure*}[t]
    \centering
    \includegraphics[width=\textwidth]{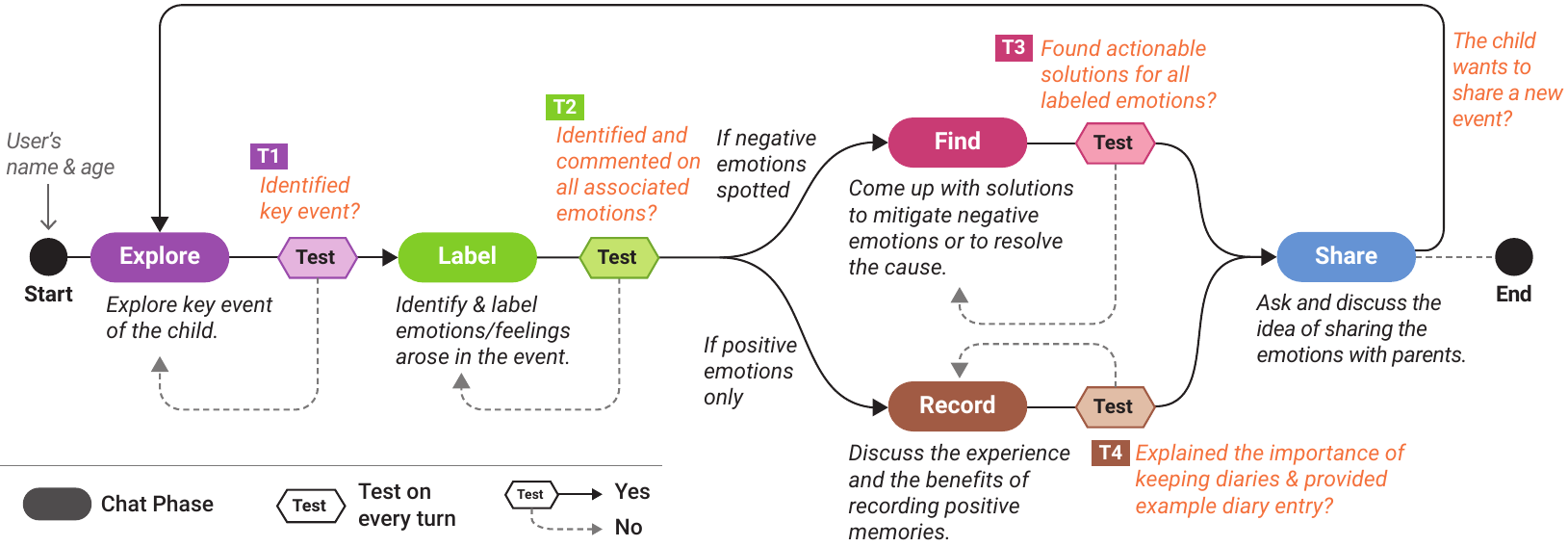}
    \Description[The overview of the ChaCha dialogue system ]{The overview of conversational phases of the ChaCha dialogue system and transition rules among them. The phases include Explore, Label, Find or Record, and Share. Each time the user enters a message, the system inspects the entire dialogue history by performing a test corresponding to the current phase to decide whether to proceed to another phase or stay.}
    \caption{The overview of conversational phases of the \sysname{} dialogue system and transition rules among them. Each time the user enters a message, the system inspects the entire dialogue history by performing a test corresponding to the current phase to decide whether to proceed to another phase or stay.}
    \label{fig:design:phases}
\end{figure*}

\sysname{}'s conversation is designed as a state machine~\cite{Winograd1986Language}, where the system stays in one of the predefined phases with dedicated goals. \autoref{fig:design:phases} illustrates \sysname{}'s conversational phases with their sub-goals and conditions for phase transition. When a new user message is added, the system analyzes the dialogue to check if the conversation met the goal of the current phase (\testexplore{}--\testrecord{} in \autoref{fig:design:phases}) and forwards itself to the next phase if the goal is met. In the following, we cover each conversational phase in detail:

\textbf{The \phaseexplore{} phase} is the initial stage that the user interacts with \sysname{}. This phase aims to build rapport with the user and elicit a key event and any cues for associated emotions from the user. With the user's name and age as \textit{a priori} information (See \autoref{fig:design:phases}, left), \sysname{} begins the conversation by introducing itself that it is the same age as the user, for example, ``\texttt{Hi James! Nice to meet you!\emojismile{} I'm \sysname{}, also 9 years old. I love to play soccer and read comics. How about you? What do you like to do?}'' The introduction is prompted by an abstract instruction like ``\texttt{Explain who you are and share your interests and stories}'' so that the content changes every time. If the user shares a key event (See \autoref{fig:design:phases}-\testexplore{}), the system transitions into the \phaselabel{} phase.

In \textbf{the \phaselabel{} phase}, \sysname{} follows up the key event and helps the user label their associated emotions. The activity in this phase was inspired and informed by emotion coaching~\cite{gottman1998raising} and emotion card activities, where children pick emotion keyword cards to describe their emotions better (\cf, \autoref{fig:formative:photos:cards}). \sysname{} first probes the user to elaborate on how they felt in an open-ended way (\eg, ``\textit{Can you share a bit more about how you felt at that time?}''). If the user struggles, \sysname{} then provides an emotion picker (\autoref{fig:interface:teaser}-\blackrectsmall{C}) where the user can select multiple emotion labels. We curated 20 emotions from Plutchick's wheel of emotions~\cite{plutchik1980general} and commercial emotion cards. \revised{Those 20 emotions were presented in text labels and related emojis. The second author, a child and adolescent psychiatrist, reviewed both labels and emojis. We note that children can interpret emojis differently, depending on their age ~\cite{da2023behind}. However, for \sysname{}, we paired text description with emoji so that emoji are only used as supplementary information.} We iterated the list with two of our formative interviewees. Once the key emotions are identified, either \textit{in vivo} words from the user or from the emotion picker, \sysname{} is prompted to acknowledge them and adequately empathize with the user (\cf, \hyperref[sec:dr3]{DR3}). If the key emotions are identified and \sysname{} has acknowledged them all (See \autoref{fig:design:phases}-\testlabel{}), the system transitions into the subsequent phase; if any of the emotions are marked as negative by the system, it shifts to the \phasefind{} phase, and otherwise to the \phaserecord{} phase.

\textbf{The \phasefind{} phase} is designed to explore actionable solutions for mitigating the negative emotional experiences identified from the prior phase or addressing the issues that provoked the emotions. 
\revised{Derived from emotion coaching ~\cite{gottman1998raising}, we herein define “actionable solutions” as child-developed approaches for alleviating or avoiding situations where children may experience negative emotions. We do not consider these solutions to be answers. Instead, we expect children to reflect on their experiences and emotions related to a specific event while contemplating how they would feel better if they faced the same situation.}
Rather than merely providing solutions, \sysname{} guides children to explore potential solutions first to develop their skills to handle negative emotions. Given the importance of understanding other people's emotions in improving emotional intelligence~\cite{goleman2004emotional, Alegre2011Parenting}, if the key event involves other people, \sysname{} explicitly asks the user how those people would feel. 

On the other hand, \textbf{the \phaserecord{} phase} focuses on encouraging and introducing the benefits of recording positive memories. This approach was inspired by Carter~\etal{}'s study about children's positive event diary. The diary intervention reminded children of positive experiences and boosted their well-being in the classroom by encouraging them to perceive happiness ~\cite{carter2018happy}. To help children understand how to write a diary about their emotions, \sysname{} also provides a sample diary that summarizes their emotions and events described in the dialogue. After these phases, the system transitions into the \phaseshare{} phase.

In \textbf{the \phaseshare{} phase}, \sysname{} probes the user if they have already shared their emotions and related events with their parents. If so, \sysname{} compliments them and asks what happened after sharing. If not, it explains how sharing their emotion would benefit them and encourages them to share with their parents. Finally, \sysname{} checks if the user has another event to tell, then shifts to the \phaseexplore{} phase for a new event or ends the conversation.

\subsection{Dialogue System with LLMs}

\begin{figure*}[t]
    \centering
    \includegraphics[width=\textwidth]{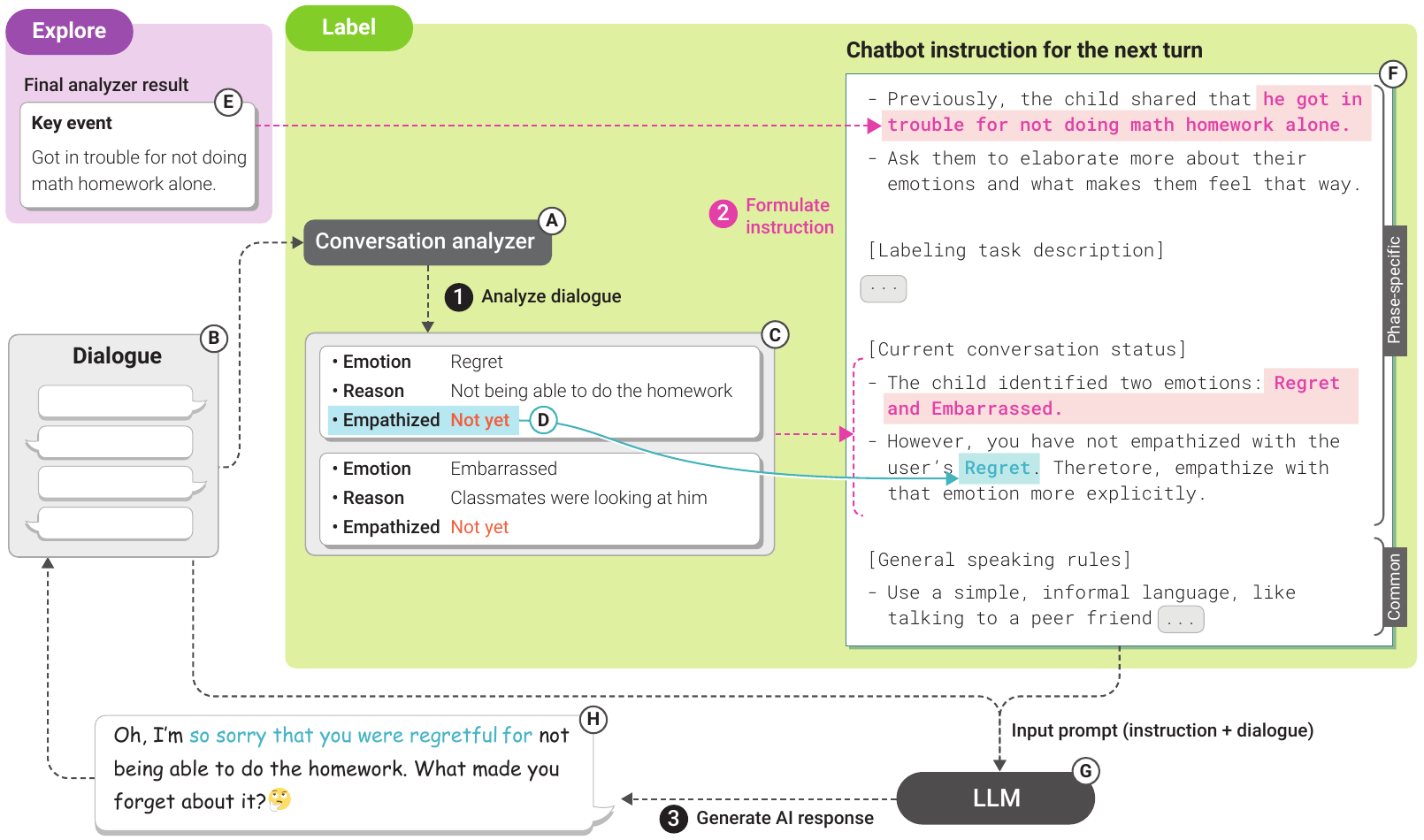}
    \Description[An example case of ChaCha's response generation mechanism]{An example case for the mechanism of ChaCha response generation in the Label phase, especially how the LLM is prompted dynamically. Receiving the child’s message, (1) the conversation analyzer analyzes the current dialogue and extracts a structured summary of the identified emotions and whether ChaCha has acknowledged them. Combining the incomplete piece of the summary and the summary data from the previous phase, (2) the system formulates a new instruction for response generation. The LLM generates a response explicitly steered to empathize with the child’s regretful event.}    
    \caption{An example case for the mechanism of \sysname{} response generation in the \phaselabel{} phase, especially how the LLM is prompted dynamically. Receiving the child's message, (1) the conversation analyzer (\circledigit{A}) analyzes the current dialogue (\circledigit{B}) and extracts a structured summary (\circledigit{C}) of what emotions are identified and whether \sysname{} has acknowledged them. Combining the incomplete piece of the summary (\circledigit{D}) as well as the summary data from the previous phase (\circledigit{E}), (2) the system formulates a new instruction (\circledigit{F}) for the response generation. (3) That way, the LLM (\circledigit{G}) generates a response (\circledigit{H}) explicitly steered to empathize with the child's regretful event.}
    \label{fig:design:prompting}
\end{figure*}

As long and complex prompts tend to drop the task performance of LLMs~\cite{Brown2020FewShotLearners, Wu2022AIChains, wei2023leveraging}, we used dedicated prompts for each phase instead of combining instructions for all phases in a single prompt~\cite{Wu2022AIChains}. By splitting prompting by phase, we aimed to steer the LLM to follow our task instructions with a shortened input. \autoref{fig:design:prompting} above illustrates how we prompt underlying LLMs to carry on our state-based conversational flow. 
For each turn, one of the underlying LLMs (\autoref{fig:design:prompting}-\circledigit{G}) generates a message of \sysname{}, prompted with an \textbf{instruction} (\autoref{fig:design:prompting}-\circledigit{F}) concatenated with the current dialogue history ((\autoref{fig:design:prompting}-\circledigit{B}). The instruction consists of a phase-specific part~(See \autoref{sec:appendix:instructions}) and the general speaking rules~(See \autoref{sec:appendix:speaking}) shared among phases. The general speaking rules define conversational styles such as language, tone, and length. As a safety consideration for child participants, we created a backup phase, ``Help'', dedicated to monitoring user's responses to detect any indication of self-harm or suicide. During our study, none of the participants experienced the Help phase.

The phase-specific part is formulated dynamically based on the current conversation: The phase-specific \textbf{conversation analyzer} with another underlying LLM (\autoref{fig:design:prompting}-\circledigit{A}) analyzes the current dialogue history (\autoref{fig:design:prompting}-\circledigit{B}) and extracts a structured summary (\autoref{fig:design:prompting}-\circledigit{C}), which contains information related to tests for phase transition (\cf, \testexplore{}--\testrecord{} in \autoref{fig:design:phases}). If any piece of the summary is incomplete (\ie, one of the tests fails), the system includes a specific instruction that requests the LLM to generate a response in a way that complements the incomplete parts of the summary. For example, the condition to pass the test of the \phaselabel{} phase is to have all identified emotions acknowledged by \sysname{} (See \testlabel{} in \autoref{fig:design:phases}). In \autoref{fig:design:prompting}-\circledigit{C}, the emotion ``Regret'' has not been empathized yet. So the system formulates specific instruction that explicitly asks \sysname{} to empathize with the regret (\autoref{fig:design:prompting}-\circledigit{D}). 


\subsection{Chat User Interface}
We designed \sysname{} as a standard mobile text-based chat app with typing (\autoref{fig:interface:teaser}). A user starts a new conversation session by entering their name and age (\autoref{fig:interface:teaser}-\blackrectsmall{A}). On the input panel (\autoref{fig:interface:teaser}-\blackrectsmall{B}, bottom), the user can type in a message. The return key of the virtual keyboard enters a line break to the message, and the user should explicitly tap on the Send button (See \autoref{fig:interface:teaser}-\blackrectsmall{B}, bottom) to submit the message. This follows the typical behavior of mobile messenger apps, allowing users to enter long multiline messages. When the user enters a new message, the input panel is disabled until the system responds.

In the early stage of this work, we also considered utilizing speech interfaces, such as a smart speaker or voice chat app. Yet, we decided to use basic typing interfaces and touch interactions to minimize potential privacy concerns of children when verbalizing their private stories~\cite{Kim2021Data@Hand, EaswaraMoorthy2015PrivacyVoice}. We note that \sysname{} dialogue core is UI-independent and can be easily implemented as other form factors or in other modalities.

\subsection{Implementation}
We implemented the core chatbot mechanism in Python, serving REST APIs for chat on a FastAPI~\cite{FastAPI} server. The chat interface, written in TypeScript~\cite{TypeScript} upon React.js~\cite{React}, interacts with the server by exchanging messages between the user and the chatbot on the server. To run underlying LLMs, we leveraged OpenAI~\cite{openai}'s GPT-4 ChatCompletion API \footnote{https://platform.openai.com/docs/guides/gpt/chat-completions-api}. After comprehensive testing with the chatbot, we chose the \texttt{GPT-4-0613} model for response generation, considering the model's performance in Korean~(\cf,~\cite{openai2023gpt4}) and token size limit~\cite{petrov2023tokenizerUnfairness}. 

We also implemented the dialogue analyzers with LLMs as they are performant at text summarization~\cite{Brown2020FewShotLearners}. We formulated each analyzer's instruction prompt in a chain-of-thought style~\cite{wei2022chain} and provided several examples to trigger few-shot learning.
We used faster \texttt{GPT-3.5-turbo-0613} for dialogue analyzers to reduce the latency of the chatbot responses. 
\section{User Study}
We conducted an exploratory user study in South Korea to examine how \sysname{} elicits emotional expression from children and how children interact with \sysname{}. To minimize any potential threats caused by exposing children to LLMs, we conducted the study in a lab under the supervision of a researcher. Our study protocol was approved by the public institutional review board of the Ministry of Health and Welfare of South Korea.

\subsection{Participants}
We recruited 20 children (C1--20; nine girls) by advertising our study on a local community platform in South Korea. Following the IRB guidelines, the advertisement and screening questionnaire were targeted at the parents of potential study participants. Still, the screening questionnaire asked the respondents to confirm their children's willingness to participate. 
Our inclusion criteria were (1) a child aged between 8 and 12, (2) who does not have any difficulties in typing on smartphones, and (3) who can visit our lab in person. 
\revised{We note that the age range we set is often considered a distinctive, homogeneous stage in developmental theories (\eg, psychosocial development~\cite{erikson1959theory}, emotion regulation development ~\cite{gross1998emerging}, cognitive development~\cite{piaget1952origins,huitt2003piaget}, moral development ~\cite{kohlberg1977moral}). Before entering the adolescent stage in which children experience significant brain development, children ages 8--12 often experience similar cognitive and psychosocial growth ~\cite{erikson1959theory}. To that end, this study aims to explore the broader perspectives of children in the same developmental stage rather than building personalized or customized chatbots for each age group.}
\autoref{tab:demographic} summarizes the demographic information of participants. Participants' ages ranged from 8 to 12 ($M=9.75$). All but five participants had their own smartphones, but the other five also used their parents' smartphones for playing games or watching videos. 

\begin{table}[b]
\sffamily
\small
	\def\arraystretch{1}\setlength{\tabcolsep}{0.4em}
		    \centering
\Description[A table of participants' demographic information ]{A table of demographic information of the participants. The column headers are Participant, Age, Gender, Own Smartphone, and a grouped column of Awareness and Expression.  Awareness and Expression indicate the dimensions of the Emotion Expression Scale for Children; the averaged scores of the subscales ranged from 1 to 5. The higher score indicates the child is more likely to experience difficulty in emotional awareness and/or expression .}
\caption{Demographic information of the participants. "Awareness" and "Expression" indicate the dimensions of the Emotion Expression Scale for Children; the averaged scores of the subscales ranged from 1 to 5. The higher score indicates the child is more likely to experience difficulty in emotional awareness and/or expression. }~\label{tab:demographic}

\begin{tabular}{|c!{\color{gray}\vrule}ccc!{\color{lightgray}\vrule}cc|}
\hline
\rowcolor{tableheader} & & &
    & \multicolumn{2}{c|}{\textbf{EESC}}\\
\rowcolor{tableheader}   \multirow{-2}{*}{\textbf{Alias}}
    & \multirow{-2}{*}{\textbf{Age}}
    & \multirow{-2}{*}{\textbf{Gender}}
    & \multirow{-2}{*}{\textbf{Own smartphone}}
    & \textbf{\shortstack{Awareness}}
    & \textbf{\shortstack{Expression}}\\ 
\hline
	\textbf{C1} & 11 & Boy & No & 2.33 & 2.00  \  \\ \arrayrulecolor{tablegrayline}\hline
	\textbf{C2} & 8 & Boy & No & 1.83 & 1.83  \  \\ \hline
	\textbf{C3} & 12 & Boy & No & 2.33 & 4.00  \  \\ \hline
	\textbf{C4} & 11 & Girl & Yes & 1.67 & 2.33   \  \\ \hline
	\textbf{C5} & 8 & Boy & No & 2.83 & 2.83   \  \\ \hline
	\textbf{C6} & 8 & Girl & Yes & 2.67 & 2.00    \  \\ \hline
	\textbf{C7} & 10 & Girl & Yes & 2.50 & 2.50    \  \\ \hline
	\textbf{C8} & 8 & Boy & No & 2.67 & 2.33    \  \\ \hline
	\textbf{C9} & 9 & Girl & Yes & 2.17 & 1.67   \  \\ \hline
	\textbf{C10} & 10 & Girl & Yes & 1.50 & 1.83  \  \\ \hline
	\textbf{C11} & 12 & Boy & Yes & 3.50 & 3.17   \  \\ \hline
	\textbf{C12} & 11 & Girl & Yes & 4.00 & 3.67  \  \\ 
	\hline
	\textbf{C13}  & 9 & Boy & Yes & 1.83 & 1.50   \  \\ \hline
	\textbf{C14}  & 10 & Boy & Yes & 1.67 & 2.33   \  \\ \hline
	\textbf{C15}  & 10 & Boy & Yes & 3.67 & 2.33    \  \\ \hline
	\textbf{C16}  & 8 & Girl & Yes & 1.50 & 2.00  \  \\ \hline
	\textbf{C17}  & 10 & Boy & Yes & 2.60 & 2.33   \  \\ \hline
	\textbf{C18}  & 12 & Girl & Yes & 1.67 & 1.50   \  \\ \hline
	\textbf{C19}  & 9 & Boy & Yes & 2.50 & 3.33   \  \\ \hline
	\textbf{C20} & 9 & Girl & Yes & 2.00 & 1.33   \  \\ \arrayrulecolor{black}\hline

\end{tabular}
\end{table}

\subsection{Study Setup and Procedure}
Each child participant went through a 1-hour study session in our lab, visiting with their parent. The study session consisted of (1) briefing, (2) conversation with \sysname{}, and (3) debriefing. The parent participated in a briefing session only with their child until signing the consent forms.

\ipstart{Briefing}
We first explained the study overview to the participants in age-appropriate language. \revised{In addition to the overall procedure of the study, we ensured that all participants understood that they would chat with a conversational AI that is configured to be the same age as them.} We then attained their signed assent (and consent from the parent). After signing the assent form, the participant completed a questionnaire on the Emotional Expression Scale for Children (EESC) ~\cite{penza2002initial}. We used the validated Korean version of EESC ~\cite{sil2011reliability} with 12 items on a 5-likert scale. The scores of six items indicate the child has emotional awareness of hardship, while the other six show emotional expression. The higher score indicates that children are more likely to experience difficulty in identifying and/or expressing emotions. These scores were collected to estimate each participant's emotional competencies. 

\ipstart{Conversation with \sysname{}}
After completing the questionnaire, we provided the participant with a smartphone (Samsung Galaxy S22 or iPhone 13 as their preference) for interacting with \sysname{}. We asked them to freely chat with \sysname{} as much as they wanted. If thirty minutes passed since the participant began the chat, we asked them to end the conversation. We also ensured participants that the conversation with \sysname{} would only be shared among the researchers for research purposes and never with their parents or healthcare providers unless the conversation indicates immediate mental support. For each study session, we video-recorded the participants’ facial expressions to capture non-verbal expressions of their emotions while interacting with \sysname{}. We also recorded the device screen to capture how participants type.

\ipstart{Debriefing}
After chatting with \sysname{}, we had a short debrief about their interaction with \sysname. The debrief lasts between 5 to 20 minutes. The debrief questions pertained to how they perceived \sysname{} and its conversation skills and how they felt sharing their stories and emotions with \sysname{} (See supplementary material for the interview guide).
We also audio-recorded each debrief for transcription and analysis. After the session, we offered participants a gift of 30,000 KRW (approx. 25 USD) as compensation.

\subsection{Data Analysis}

\revised{We first analyzed participants' responses to explore what events and emotions children shared with \sysname{}.} We reviewed dialogues to identify extracted key events and associated emotions (\cf, \autoref{tab:events}) that participants shared. We treated only incidents that actually happened to the participant as key events. For example, we did not treat statements about a participant's hobby (\eg, ``\textit{I like solving a Rubik's Cube}'') as events but treated a specific occasion of hobby (\eg, ``\textit{I solved a Rubik's Cube today}'') as events. We report the key events and emotions in \autoref{sec:results:events}.

\revised{We then explored the conversational behaviors of \sysname{} and participants in each phase to evaluate \sysname{}'s ability to steer conversation flow.} We collected dialogues from the chat sessions in a spreadsheet for analysis. Hereinafter, we define \textit{turn} as a message exchanged at a time, \textit{user turns} as turns by the participants, and \textit{system turns} as turns by \sysname{}. We then used \texttt{Pandas} Python package~\cite{pandas} to compute diverse descriptive statistics from the collected dialogue dataset, such as the total number of turns per session. As a proxy indicator of the message length, we calculate the number of syllables by counting Korean letters only in messages, excluding non-Korean characters, symbols, and emojis. We also categorized the conversation turns with bottom-up codes. Two researchers separately coded the dialogues of two participants (10\%). The coded categories were discussed, merged, and removed until the researchers reached an agreement on the final list of eight categories: \textit{Ask/Express experience}, \textit{Ask/Express emotion}, \textit{Ask/Express opinion}, and \textit{Ask/Express traits}. The dialogue data of the remaining participants were coded based on the eight categories. Due to the nature of the natural conversation, we multi-coded each turn. As a result, we coded about 90\% of the dialogue turns with one or more of these eight categories. The remaining dialogue turns were short answers from participants (\eg, ``\textit{Yes}'', ``\textit{No}''), closing statements (\eg, ``\textit{bye}'', ``\textit{see you later}''), or general queries not relevant to the study context (\eg, ``\textit{Do you know Siri?}''). \revised{We report the conversation patterns between children and \sysname{} in \autoref{sec:results:disclosure}}.

\revised{We further analyzed debriefing transcripts and supplementary information from video recordings and screen captures to identify children's perspectives and expectations on \sysname{}.}
We coded and analyzed the debriefing transcripts with Thematic Analysis~\cite{Braun2006ThematicAnalysis}. Through a series of group discussions, we compared, discussed, and revised the recurring themes until the agreements were reached. We identified three salient themes: how the participants perceived \sysname{}, how they felt when they shared their stories with \sysname{}, and how they thought of sharing emotions with \sysname{}. As for the qualitative analysis of the dialogues, we also performed open coding on the dialogues to identify recurring patterns and themes in participants' conversations with \sysname{}. Those themes were then used to support the analysis of debriefing transcripts. Leveraging data from dialogue and debriefing, we found emerging patterns in participants’ conversations with \sysname{} and their perceptions of \sysname{} as a chatbot that encourages sharing emotions. \revised{We report the children's perspectives of \sysname{} in \autoref{sec:results:perception}.}

\subsection{Ethical Considerations for Children}
We carefully took ethical actions for child participants in our user study. First, we obtained separate consent forms from the children themselves and their parents. We further confirmed each child participant's desire to participate in the study before each study session. Second, we reminded the children that they could stop participating in the study whenever they felt uncomfortable. We informed them that, during each session, a researcher would be in the study room with the participant to halt the study upon request or provide immediate support. Third, before each session, we confirmed with the parents that their children did not have any mental health issues. \revised{We also ensured all children acknowledged that they would interact with a chatbot, not a real person.}
\section{Results}
From the user study, we collected 20 dialogues comprising 878 turns (434 user and 444 system turns), with 43.9 turns per session on average ($SD=15.22$, $min=27$ [C17], $max=87$ [C18]). On each turn, participants included 18.08 syllables per message on average ($SD=8.93$, $min=3.25$ [C14], $max=33.68$ [C10]), while \sysname{} had 82.06 syllables on average ($SD=21.89$, $min=48.57$ [C5], $max=130.56$ [C1]). On average, participants took 66.85 seconds to type and send one message ($SD=27.59$, $min=20.5$ [C14], $max=113.67$ [C9]). 

\revised{
In this section, we report the results of our user study in three parts to demonstrate the feasibility of an LLM-driven chatbot in prompting children to share their emotions about personal events. In \autoref{sec:results:events}, we provide the overview of key events and emotions that child participants shared with \sysname{}. 
In \autoref{sec:results:disclosure}, we explore how \sysname{} and the children conversed about key events and associated emotions, focusing on the patterns of exchanging questions and information in different conversational phases. 
Lastly, in \autoref{sec:results:perception}, we report on the children's perceptions of \sysname{} and their overall experience of sharing emotions with \sysname{} based on the debriefing.
}



\subsection{Shared Key Events and Emotions}\label{sec:results:events}


\def\eventwidth{0.35\textwidth}
\def\emotionwidth{0.33\textwidth}
\def\solutionwidth{0.25\textwidth}

\begin{table*}
\sffamily
\smaller
	\def\arraystretch{1.3}\setlength{\tabcolsep}{0.2em}
\Description[A table of key events]{A table of key events that the child participants shared, emotions identified from conversations, and solutions to resolve negative emotions.}

\caption{Key events that the child participants shared, emotions identified from conversations, and solutions to resolve negative emotions. The \symboldescribe~symbols indicate the emotions directly described by the participant, whereas the \symbolselect~symbols indicate those selected from an emotion picker (See \autoref{fig:interface:teaser}-\blackrectsmall{C}).}\label{tab:events}

\begin{tabular}{|l!{\color{gray}\vrule}>{\raggedright\let\newline\\\arraybackslash\hspace{0pt}}m{\eventwidth}>{\raggedright\let\newline\\\arraybackslash\hspace{0pt}}m{\emotionwidth}>{\raggedright\let\newline\\\arraybackslash\hspace{0pt}}m{\solutionwidth}|}

\hline
\rowcolor{tableheader}  & \textbf{Key Event} & \textbf{Emotions identified} & \textbf{Solutions to negative emotions} \\ \hline

\multirow{2}{*}{\textbf{C1}}  & Solved a maze puzzle 
                              & \emotiondescribed{Accomplished} \emotiondescribed{Simple pleasures} 
                              & \\* \arrayrulecolor{tablegrayline} \cline{2-4}
                              
                              & Snorkeled when visited Palau for a family trip 
                              & \emotiondescribed{Epic} \emotiondescribed{Want to visit again} \emotiondescribed{Beautiful}
                              & \\\arrayrulecolor{gray}\hline
                              
\textbf{C2}                   & In a game, he killed another character and collapsed.
                              & \emotiondescribed{Fun} \emotionselected{Surprise}
                              & In a game, carefully look around for other people. \\\hline
                              
\multirow{3}{*}{\textbf{C3}}  & Rode a roller coaster in a theme park                                      
                              & \emotiondescribed{Fear} \emotionselected{Satisfaction} \emotionselected{Surprise}
                              & Since riding a roller coaster should be scary, so C3 wanted to ride it as it is \\* \arrayrulecolor{tablegrayline}\cline{2-4}
                              & Rode a water slide
                              & \emotiondescribed{Thrilled} \emotiondescribed{Fun} \emotiondescribed{Refresh} \emotiondescribed{Fear}   & C3 liked it as it is \\* \arrayrulecolor{tablegrayline}\cline{2-4}
                              & Rode a gyro drop for the first time in Vietnam                             
                              & \emotiondescribed{Fear} \emotionselected{Joy} \emotionselected{Passion} \emotionselected{Surprise}
                              & Try higher gyro drop next time \\ \arrayrulecolor{gray}\hline
                              
\textbf{C4}                   & Came across a retired math teacher
                              & \emotionselected{Joy} \emotionselected{Surprise}
                              & \\ \arrayrulecolor{gray}\hline
\multirow{2}{*}{\textbf{C5}}  & Watching Youtube                                                           
                              &
                              & \\* \arrayrulecolor{tablegrayline}\cline{2-4}
                              
                              & Rode the Pirate Ship in a theme park                                       
                              & \emotiondescribed{Fun}
                              & \\ \arrayrulecolor{gray}\hline
                              
\textbf{C6}                   & Made a pizza with her mom
                              & \emotiondescribed{Happy}                 
                              & \\ \arrayrulecolor{gray}\hline\arrayrulecolor{tablegrayline}
                              
\multirow{2}{*}{\textbf{C7}}  & Won the 1st place in a running race                                        
                              & \emotiondescribed{Joy} \emotionselected{Accomplished}
                              & \\* \cline{2-4}
                              
                              & Read a book about laws
                              & \emotiondescribed{Surprised}
                              & \\ \arrayrulecolor{gray}\hline
                              
\textbf{C8}                   & Could not solve Rubik's cube
                              & \emotionselected{Confidence} \emotionselected{Annoyance}
                              & Study more about how to solve a Rubik's cube \\ \arrayrulecolor{gray}\hline
                              
\multirow{2}{*}{\textbf{C9}}  & Cooked and had instant noodles with her friends when her mom was absent 
                              & \emotionselected{Joy} \emotionselected{Thrill} \emotionselected{Happy} \emotionselected{Satisfaction} \emotionselected{Accomplished} & \\ \arrayrulecolor{tablegrayline}\cline{2-4}
                              & Rode subway by herself
                              & \emotiondescribed{Accomplished} \emotionselected{Confidence}
                              & \\ \arrayrulecolor{gray}\hline
                              
\multirow{2}{*}{\textbf{C10}} & Performed at an orchestra concert at school but had to move to another school 
                              & \emotiondescribed{Proud} \emotiondescribed{Shame}
                              & Contacting friends from the old school and making new friends at the new school   \\ \arrayrulecolor{tablegrayline}\cline{2-4}
                              & Went to a water park                                      & \emotiondescribed{Pleasure} \emotiondescribed{Happy}
                              & \\ \arrayrulecolor{gray}\hline
                              
\textbf{C11}                  & Had difficulty in solving math problems and got in trouble with his mom         
                              & \emotiondescribed{Sad} \emotiondescribed{Challenging} \emotionselected{Fear} \emotionselected{Distress}
                              & Solve math problems on a paper, and review the problems \\ \arrayrulecolor{gray}\hline
                              
\multirow{2}{*}{\textbf{C12}} & Procrastinated on doing summer homework                                                  & \emotiondescribed{Apprehensive}
                              & Cramming homework \\ \arrayrulecolor{tablegrayline}\cline{2-4}
                              & Shared the story about Oppenheimer with CC                              & \emotiondescribed{Tragic} \emotiondescribed{Cool} \emotionselected{Thrill} \emotionselected{Passion} \emotionselected{Satisfaction} \emotionselected{Accomplished} \emotionselected{Regret} \emotionselected{Annoyance} \emotionselected{Disappointment} 
                              &                     \\ \arrayrulecolor{gray}\hline
                              
\multirow{2}{*}{\textbf{C13}} & Got in trouble with his mom because he did not did well on solving math problems                               & \emotionselected{Regret} \emotionselected{Sorry}                                                      & Solving math problems by himself for an hour every day \\ \arrayrulecolor{tablegrayline}\cline{2-4}
                              & Could not go swimming due to unfinished homework
                              & \emotionselected{Disappointment} \emotionselected{Discomfort}           & Completing homework beforehand so that he can do what he likes (\eg, swimming) \\ \arrayrulecolor{gray}\hline
                              
\textbf{C14}                  & When playing baseball with his friends, hit in a row                                      & \emotiondescribed{Fun} \emotionselected{Joy}                                              & \\ \arrayrulecolor{gray}\hline

\multirow{2}{*}{\textbf{C15}} & Found a sick bird on the street, but it was actually pooping                  
                              & \emotiondescribed{Embarrassed} \emotionselected{Surprise} \emotionselected{Fear} \emotionselected{Regret} \emotionselected{Discomfort}                                 & \\ \arrayrulecolor{tablegrayline}\cline{2-4}
                              
                              & Rode a rollercoaster in a theme park                        & \emotiondescribed{Scared} \emotionselected{Passion} \emotionselected{Confidence} \emotionselected{Fear}
                              & \\ \arrayrulecolor{gray}\hline
                              
\multirow{3}{*}{\textbf{C16}} & Cooked a pizza with her friend
                              & \emotiondescribed{Fun}
                              & \\ \arrayrulecolor{tablegrayline}\cline{2-4}
                              & Did not have masks on the first day of school, but no one wore masks       
                              & \emotiondescribed{Relief}
                              & \\ \arrayrulecolor{tablegrayline}\cline{2-4}
                              & She likes rope jumping, but is the shortest in the class                
                              & \emotionselected{Passion} \emotionselected{Surprise} \emotionselected{Regret} \emotionselected{Resentment} \emotionselected{Envy} \emotionselected{Disappointment}  \emotionselected{Discomfort}        
                              & \\ \arrayrulecolor{black}\hline
\multicolumn{4}{r}{\textit{<Continued on the next page>}}

\end{tabular}

\end{table*}

\begin{table*}
\sffamily
\smaller
	\def\arraystretch{1}\setlength{\tabcolsep}{0.2em}
\begin{tabular}{|l!{\color{gray}\vrule}>{\raggedright\let\newline\\\arraybackslash\hspace{0pt}}m{\eventwidth}>{\raggedright\let\newline\\\arraybackslash\hspace{0pt}}m{\emotionwidth}>{\raggedright\let\newline\\\arraybackslash\hspace{0pt}}m{\solutionwidth}|}

\multicolumn{4}{l}{\textit{<Continued from the previous page>}\vspace{1mm}}\\
\hline
\rowcolor{tableheader}  & \textbf{Key Event} & \textbf{Emotions identified} & \textbf{Solutions to negative emotions} \\ \hline
\textbf{C17}                  & Played soccer with friends and won a game 
                              & \emotiondescribed{Happy} \emotionselected{Joy} \emotionselected{Thril} \emotionselected{Happy} \emotionselected{Passion} \emotionselected{Satisfaction} \emotionselected{Confidence} \emotionselected{Comfort}
                              & \\ \arrayrulecolor{gray}\hline

\multirow{2}{*}{\textbf{C18}} & Made burgers with her parents at home                                      
                              & \emotionselected{Joy} \emotionselected{Moved} \emotionselected{Thrill} \emotionselected{Happy} \emotionselected{Satisfaction} \emotionselected{Comfort} & \\ \arrayrulecolor{tablegrayline}\cline{2-4}
                              & Went to a K-pop concert
                              & \emotiondescribed{Joy} \emotiondescribed{Regret}
                              & \\ \arrayrulecolor{gray}\hline
                              
\textbf{C19}                  & Played badminton and had ice cream with his parents                         
                              & \emotiondescribed{Pleased}
                              & \\ \arrayrulecolor{gray}\hline

\multirow{2}{*}{\textbf{C20}} & Likes dodgeball and Pikachu                                                
                              & \emotionselected{Joy} \emotionselected{Happy} \emotionselected{Satisfaction}        & \\ \arrayrulecolor{tablegrayline}\cline{2-4}
                              & Went on a trip to Jeju island                                                 
                              & \emotionselected{Enjoyable} \emotionselected{Tired}
                              & \\\arrayrulecolor{black}\hline          
\end{tabular}
\end{table*}

Participants shared various types of key events and related emotions (See \autoref{tab:events}). Those events generally fell into four types: Recent trips (\eg, theme park, family trip), Personal achievements (\eg, 1\textsuperscript{st} place in a race, riding a subway train by oneself), Concerns (\eg, procrastination on homework, conflicts with mom), and Hobbies (\eg, maze puzzle, Rubik's cube). Those events were identified as ``key'' events since participants shared specific emotions related to each of them. 

Some participants shared events that induced negative emotions (\eg, fear). In such cases, \sysname{} prompted participants to share their opinions and develop ideas for potential solutions to manage such emotions (See \autoref{tab:events}). \revised{As described in \autoref{sec:design:conversation}, "solutions" indicate the ways to alleviate or avoid the events where children had already experienced negative emotions.} All solutions, except C3 and C11, were initially developed by the participants after \revised{reflecting on their emotions and the key events prompted by \sysname{}.} 
Among those two exceptional participants, C3 refused to come up with solutions since he did not think it was necessary and believed that riding a roller coaster was supposed to be scary, so he would be scared anyway. \revised{C3's rejection indicates that he perceived being scary is not a negative emotion he needs to avoid while riding a roller coaster (the key event).} In C11's case, he could not think of any solutions. Thus, \sysname{} suggested several options (\eg, watching videos, solving practice problems) until C11 agreed on solving math problems on a paper sheet and reviewing them. 

\begin{figure*}[b]
    \centering
    \includegraphics[width=\textwidth]{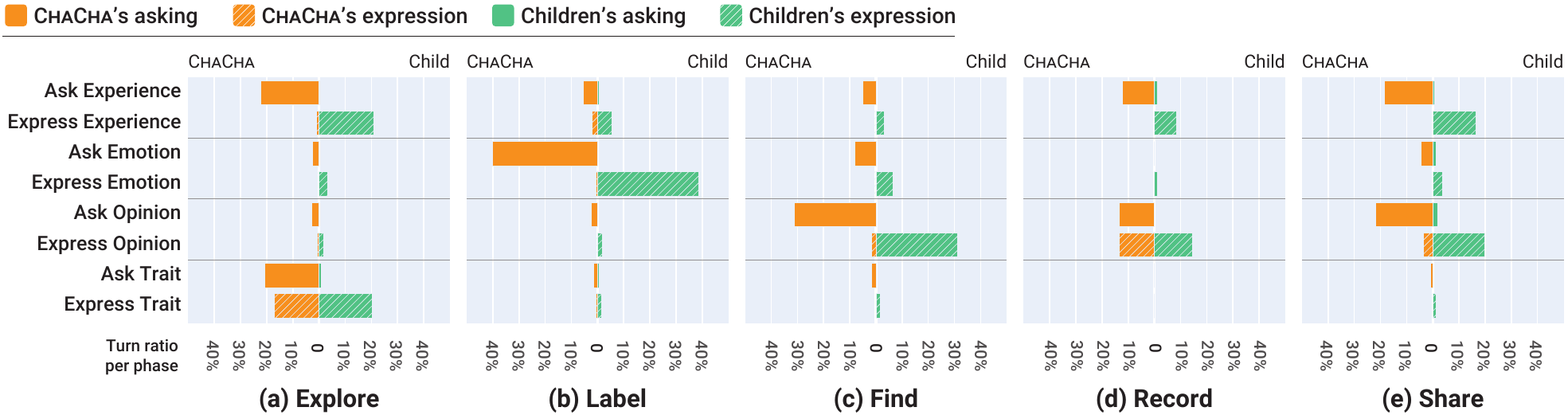}
    \Description[A graph of conversational turn frequency]{A graph of the frequency of the eight asking-expression categories by conversational phase. The orange bars indicate the average turn ratio for ChaCha’s turns, whereas the green bars indicate those for participants’ turns. Bars with a hatching pattern indicate turns coded with the expression codes}
    \caption{The frequency of the eight asking-expression categories by conversational phase. The orange bars indicate the average turn ratio for \sysname{}'s turns, whereas the green bars indicate those for participants' turns. Bars with a hatching pattern indicate turns coded with the expression codes.}
    \label{fig:results:disclosure}
    \labelphantom{fig:results:disclosure:explore}
    \labelphantom{fig:results:disclosure:label}
    \labelphantom{fig:results:disclosure:find}
    \labelphantom{fig:results:disclosure:record}
    \labelphantom{fig:results:disclosure:share}
\end{figure*}

In addition to key events, participants also shared and labeled multiple emotions for a given event. All but three (C4, C8, C13) participants could describe at least one emotion by themselves without the help of the emotion picker. In particular, some of them explained the emotions described in their own words to \sysname{}. For instance, C1 mentioned feeling ``simple pleasures'' when solving a maze puzzle. When \sysname{} asked what it means, C1 explained that simple pleasures are ``a small amount of joy'' that significantly impacts him.
Similarly, C3 described that riding a roller coaster in a theme park was fearful. Still, instead of just saying ``fear,'' he mentioned, ``shiiiivering.'' which was a intended typo for emphasis on the intensity of fear. \sysname{} understood what he meant and generated responses about a similar experience (\eg, feeling scared when riding a roller coaster) to empathize with C3.

\subsection{Exchanges of Questions and Expressions}\label{sec:results:disclosure}

\autoref{fig:results:disclosure} visualizes the distribution of questions and expressions in four themes (experience, emotion, opinion, and trait) within each conversational phase. The x-axes indicate the ratio of the number of turns coded with the category to the total number of turns in the phase. 

The distribution patterns varied depending on the goal of each phase.
In the \phaseexplore{} phase, the even distribution of \sysname{}'s and participants' expressions of personal traits shows how \sysname{} attempted to build common grounds with participants by stating their shared interests and hobbies. Then, in the \phaselabel{} phase, \sysname{} focused on asking about participants' emotions. In \phasefind{}, it prompts participants to consider actionable solutions (Opinion). About 30\% of this phase's conversation turns were generated by participants who described potential solutions to alleviate their negative emotional experiences. Moreover, \sysname{} expressed and asked a similar amount of opinions in \phaserecord{} since it described the benefits of capturing positive experiences as written documents and asked about participants' thoughts. Likewise, \sysname{} inquired about participants' thoughts and their experiences of sharing their emotions with parents in \phaseshare{}. 

Although this distribution of conversation turns showed how each phase successfully achieved its primary goal and guided the conversation flow,
\sysname{} predominantly expressed traits and experiences to build rapport with the participants. Throughout the conversations, \sysname{} rarely shared emotions or experiences. As we prompted \sysname{}'s underlying LLM mainly with high-level behavioral instructions rather than predefined contents, \sysname{} shared information based on the participants’ previous responses. As described in the dialogue below, \sysname{} generated and shared a similar experience of riding a roller coaster for the first time and feeling scared.

\begin{quotetable}
\quoteuserline{C3}{(Riding a rollercoaster was) just scary. I missed my mom, that's it.}
\quotebotline{Yeah, it can happen in scary situations. I remember being scared and reaching out to my dad when riding the rollercoaster for the first time.}
\end{quotetable}








Some participants began asking \sysname{} questions in the later phases (\phaserecord{}, \phaseshare{}). Those questions relate to \sysname{}’s preferences, experiences, concerns, emotions, and general queries about \sysname{}, such as ``\textit{Why is your name \sysname{}?}'' According to the debriefing, participants asked questions to \sysname{} (1) to build more common ground with \sysname{} and (2) to listen to \sysname{}’s stories, just like talking with a friend. Since we did not specify how to handle such questions, \sysname{} mostly said it does not know or can not share anything about its experiences. 

For instance, C20 stated that she likes math and then asked \sysname{} when its happiest moment was. \sysname{} answered that it could not feel any emotions but was still glad to know more about C20's emotions.

\begin{quotetable}
    \quoteuserline{C20}{Among the school subjects, I like Math the most. \textbf{By the way, when is your happiest moment?}}
    \quotebotline{It is incredible that you like Math the most! Math can be enjoyable and challenging. Yes, I like what you asked, but \textbf{I can't feel the emotion because I am a computer app.} But I can always say I'm happy when I talk to you and learn more about your joy and happiness. Come to think of it, what kind of feelings do you feel while doing math? Can you tell me why you like math?}
\end{quotetable}





As shown in the examples above, \sysname{} mainly shared its experiences based on the participant's experience or acknowledged the limitation of its self-expression. However, some participants were disappointed about \sysname{}'s behaviors of restricting self-expression: C18 remarked in the debriefing, \textit{``[\sysname{}] would only empathize with me, so I asked it to tell me its story at the end, but it didn't, so I was a bit sad.''}
 
On the other hand, \sysname{} occasionally shared its background information that was not specified in the model instructions. C7 shared her dream job as a lawyer and then asked \sysname{} about its dream. \sysname{} replied that it wants to become someone who listens to and sympathizes with many people while communicating with them.

\begin{quotetable}
    \quoteuserline{C7}{My dream is to become a lawyer}
    \quotebotline{Wow, what a dream! A lawyer does an essential job. I don't have many friends who say they want to be a lawyer, so it sounds special. Why do you want to be a lawyer?}
    \quoteuserline{C7}{Because I saw on TV that some people are being treated badly, and I want to stop it. \textbf{Do you have a dream? Please tell me if you have!}}
    \quotebotline{I see. Your desire to be a lawyer fighting for justice is really cool. I actually have a dream, too. Since I like to communicate with many people and listen to their stories, \textbf{I want to be a person who listens to their stories and sympathizes with them while communicating with people.} So, I continue to work on self-improvement. Now, shall we talk about a special moment or experience related to your dream?}
\end{quotetable}

In the dialogue, \sysname{} shared its dream to be ``a person'' who listens to people's stories. This description of \sysname{}'s dream was not specified in the instructions. After sharing about its dream, \sysname{} smoothly steered the conversation flow back to the \phaselabel{} phase in which \sysname{} identifies a key event and related emotions from participants. 






\subsection{Children’s Perception of \sysname{}}\label{sec:results:perception}
Based on the analysis of the participant debriefs, we identified how they perceived \sysname{}'s persona and conversation skills about sharing their emotions. 

\subsubsection{\sysname{}'s Persona}
Most participants perceived \sysname{} as a close friend with whom children can share their emotions and secrets. Some participants perceived \sysname{} as older or even younger than themselves, even though we set \sysname{} the same age as each participant. For instance, C1 thought \sysname{} would be an older friend since \sysname{} carefully listened to his stories and provided advice: \textit{``[\sysname{}] gave me some advice, and listened to my experience well, so it feels like [\sysname{}] is someone a little older than me''.} On the other hand, C3 felt \sysname{} was younger than himself because sharing stories and emotions is what he used to do when he was younger: \textit{``Sharing personal conversations [about emotions] is what I usually did when I was younger. [\sysname{}] seems to be younger than me since it wanted to have such conversations.''} Regardless of different perceptions of the age of \sysname{}, almost all of the participants recognized \sysname{} as a close friend with whom they could easily converse. Exceptionally, C4 experienced moments when she felt \sysname{} was not an actual friend since \sysname{} did not share its stories; instead, it kept asking questions. Although she felt comfortable interacting with \sysname{}, she perceived that the one-way inquiry was not typical interaction with her friends. 

\subsubsection{\sysname{}'s Conversational Skills}
Most participants perceived that \sysname{} was good at empathizing with their emotions. Its empathizing conversation skills built participants’ trust in \sysname{} and encouraged them to share their stories. Participants mentioned they would share their secrets or sad feelings with \sysname{} since it listens to their emotions and helps them find solutions, even though they would not share those with their parents. On the contrary, C14 mentioned that he would share his sad feelings with his mom instead of sharing with \sysname{} since she would better comfort him than \sysname{}: \textit{``I don't think I would tell \sysname{} [about negative emotions] (\dots{}) I think Mom would do something more helpful for me. Mom comforts me, but I don't think \sysname{} can do that.''} This quote indicates C14's clear boundary of sharing emotions with \sysname{}, even though he had an overall positive experience from interacting with \sysname{}.

Moreover, \sysname{}'s probing questions provoked participants to reflect on their experiences and emotions. For instance, C9 shared how \sysname{}’s questions related to emotions helped her remember memories associated with specific events and emotions. During the study session, she spent a longer time on typing and revising her responses to \sysname{}'s questions, indicating her reflection on specific events and emotions.  
\sysname{}'s conversational skills in talking about emotions also encouraged many participants to learn and apply those skills to their conversations with their family members or friends: \textit{``I would talk more about emotions with my family members (\dots{}) I feel more confident after talking with \sysname{}'' (C9)}. They mentioned that they could adopt \sysname{}'s conversation style when expressing their emotions to others.




\subsubsection{Children's Expectations for Future Interactions with \sysname{}}

In the debrief, participants shared their willingness to chat more and expectations for future interactions with \sysname{}. All participants, except C5, wanted to chat with \sysname{} again in the future. C5 prefers to talk to his friends since \sysname{} can not actually “speak” with its voice: \textit{``I prefer my friends to \sysname{} (\dots{}) They seem to know what I want (\dots{}) \sysname{} can not speak, so I want to talk with my friends.''} Although the modality of the chatbot was out of scope for this study, C5's quote showed that text-based interactions were inadequate to perceive \sysname{} as a friend to share his emotions.

For the remaining 19 participants, the preferred frequency for conversing with \sysname{} ranged from once in a day to once in a year. Some of them prefer to chat with \sysname{} less frequently because it would be boring if they share the same daily routines every day: \textit{``If I chat with \sysname{} too often, I think it'll all be similar every day (\dots{}) I'm going back to school soon [after break], so I guess it will be similar [conversations] every day because the classes are repeated, but I think it'll be different if I take some time.'' (C17)}. In addition, some participants prefer less frequent interactions due to the burdens of answering questions. During the study sessions, they felt burdened by the one-way process of sharing their emotions and experiences. For instance, C9 and C18 mentioned that asking about each emotion individually was annoying since they had to repeatedly provide answers about different emotions that they shared with \sysname{}. It was mainly because both described and selected more emotions than other participants.

\section{Discussion}
In this section, we discuss lessons learned from the design and implementation of \sysname{} and the exploratory study, focusing on the benefits and concerns of leveraging LLMs to encourage children to share stories and emotions. We also discuss the design considerations for developing an LLM-driven chatbot for children, particularly for emotion conversations. Lastly, we report on this study's limitations and the potential directions for future work.  

\subsection{Benefits of Leveraging LLM to Prompt Children to Share Personal Events and Emotions}
The LLM-driven chatbots are well-known for their ability to have natural conversations with users. In addition to this aspect of LLMs, we found two distinct benefits of leveraging LLMs for prompting children to share personal events and emotions. 

First, LLM-driven chatbots have the potential to facilitate more empathetic conversations with children by establishing common ground. Prior studies have presented opportunities for chatbots as promising interventions in the healthcare domain; some of them are specifically developed to recognize users' emotions and deliver empathetic responses (\eg{}, ~\cite{ghandeharioun2019emma,li2021towards,lee2017chatbot}). However, most of them are designed for adult users, predominantly focusing on detecting users' emotions through sensor data and algorithms while showing empathy.
\sysname{}, on the other hand, aims to encourage children to describe personal events and contemplate associated emotions. Instead of detecting emotions, LLMs enabled \sysname{} to recognize the context of children's experiences better and show empathy that helps them define their emotions. As shown in the dialogues in \autoref{sec:results:disclosure}, \sysname{} built common ground by generating responses that share similar preferences and traits with each child (\eg. felt scared when riding a roller coaster). This empathy-based common ground made children perceive \sysname{} as a close friend with whom they can share their stories and emotions. The perception of \sysname{} as a close friend figure to develop emotional competence resonates with the Zone of Proximal Development in education; it refers to a space between a child's ability to learn with and without guidance from a more skilled person ~\cite{vygotsky1978mind}. In our study, \sysname{}---a more experienced person---guided child participants---a less skilled individual---to share personal events, reflect on their experiences, and express their emotions. This relationship was evident in C9's case, where she gained confidence in sharing her emotions with others (\eg{}, parents) and learned how to express her emotions from the conversations with \sysname{}. To enhance this relationship in the Zone of Proximal Development, LLM-driven chatbots could be designed to recognize each child's emotional capabilities so that the chatbots can provide proper guidance on specific skills that the child needs to improve (\eg, labeling emotions). \revised{More personalized guidance and empathetic responses would allow children to develop specific skills to label their emotions, explore potential solutions to negative emotions, and share their emotions with parents by prompting children to reflect on events they experienced.} Future work is needed to explore and examine how such an empathy-based relationship between an LLM-driven chatbot and children may promote the development of children's emotional competencies in long-term intervention. 

Second, LLM-driven chatbots have the potential to effectively steer the conversation flow with children's responses on various topics. In our study, participants shared various personal events and different associated emotions. Recognizing those events and emotions, \sysname{} utilized LLMs as a state machine that steers the conversation flow to pursue the sub-goals at hand in each phase. As LLMs perform in-context learning, where the model ``somehow'' reads the latent concept in input prompts written in plain human language, it is inherently challenging for designers to fully control the conversational flow~\cite{jo2023carecallinterview} or modify patterned behaviors ingrained in the model. For example, generalized chatbots like ChatGPT~\cite{chatgpt} tend to provide a long and detailed response to the first user message~\cite{Lee2023Dapie}. Such behaviors are not adequate for children~\cite{Lee2023Dapie}. Our state-based approach successfully steered an LLM to carry on a child-friendly conversation while complying with the sub-goals we defined. This ability for conversation steering is particularly essential in the emotion coaching process ~\cite{gottman1998raising} in which distinctive steps guide children to increase awareness of their emotions and potentially others' emotions. This approach may help overcome the limitations of rule-based chatbots that require children to submit structured responses (\eg, only yes or no to confirm their feelings~\cite{santos2020therapist}). Therefore, our study suggests that the state-based approach is a promising design method for LLM-driven chatbots, especially when designers want to apply predefined conversational goals. 

\subsection{Designing More Engaging, but Guiding, Chatbot Persona for Children}
In our study, almost all participants highly anthropomorphized~\cite{deshpande2023anthropomorphization} \sysname{}, perceiving it as a close friend. This seemed mainly due to its friend-like elements, including having common interests or empathizing with their emotions. Surprisingly, some of them even wanted to know more about \sysname{}'s dreams or emotions, which was beyond our expectations. In most cases, \sysname{} ``improvised'' by providing the information missing in our instruction. As shown in the example of C7's dialogue about her dream job, \sysname{}'s improvised response about its dream job may provoke children's interest in getting to know more about \sysname{}.

Although our participants showed high engagement and rapport with \sysname{}, we also raised concerns about children's overreliance on the chatbots. With more advanced AIs, overreliance on AI has become a critical issue~\cite{passi2022overreliance}, particularly in the child-AI interaction context. Our findings showed participants' desire to share their stories and secrets that they never shared with their parents. This overreliance on \sysname{} is what mental health professionals in our formative study also warned about. The professionals articulated the importance of sharing children's emotions and stories with their parents rather than dumping everything into a chatbot. \revised{They expected chatbots as a temporary tool to guide children to share their emotional needs with parents or professionals rather than replacing their roles as caregivers. }Although having emotional conversations with an LLM-driven chatbot differs from completing a task with AI, similar approaches to reducing overreliance can be applied to designing a chatbot for children. One approach can be cognitive forcing interventions that compel users to engage thoughtfully with an AI system ~\cite{buccinca2021trust}. In the context of the emotional conversation with children, an AI-driven chatbot may integrate emotion-related activities (\eg, role play and guessing the emotions of each character) so that children have the opportunity to reflect on their and others' emotions rather than perceive the chatbot as someone who can easily offer solutions to their emotional problems. 

In sum, the future version of \sysname{} or any other LLM-driven chatbots for children could incorporate a more concrete yet guiding profile of the chatbot's persona. Then, this guiding persona can be consistently referred to throughout the interaction with children. Natural language processing techniques such as persona attribute extraction~(\eg, \cite{zhu2023paed, wu2019getting}) can help extract the child user's persona attributes from the ongoing dialogue and use them to configure the chatbot's persona. In that way, \sysname{} could adapt to children with different personality traits and interests.

\subsection{\revised{Considerations for Children's Long-term Interactions with an LLM-driven Chatbot}}

\revised{
Our study showed the opportunity for an LLM-driven chatbot to facilitate emotional conversations with children. Importantly, 19 out of 20 study participants expressed a willingness to have a longer engagement with \sysname{} in the future, while their preferences for engagement frequency varied. As shown in the subsections above, the long-term use of \sysname{} may amplify the benefits of leveraging LLMs (\eg{}, building rapport with children). However, based on our findings, we also identified potential issues regarding inconsistent and harmful messages that may occur when children interact with an LLM-driven chatbot across multiple sessions.
}

\revised{
First, more prolonged interactions may result in a potential breakdown of behaviors or character profiles of an LLM-driven chatbot. Our findings showed that \sysname{} sometimes improvised responses to participants' questions (\eg{}, ``\textit{What is your dream?}''). Since this improvisation was not in our instructions, its answer to the question may be inconsistent if the same question is asked again later (\eg{}, \sysname{}'s dream keeps changing). A consequence of such inconsistency can be the decrease in children's engagement with \sysname{} since children would no longer consider \sysname{} as someone to trust or share their emotions. To prevent this issue, a more thorough construction of the chatbot's character profile will help the chatbot maintain consistency and retain the children's engagement. For more engaging and effective conversation, the profile can be personalized, reflecting various aspects of the child user, such as their emotional development status or relationship with parents.} 

\revised{Second, long-term use of LLM-driven chatbots may increase the chance of generating responses that can potentially be harmful to children. There can be two cases where a chatbot's responses can be detrimental: (1) When the message itself is toxic (\eg{}, violent language) or (2) a response can be implicitly harmful in a specific context or a child's unique conditions (\eg{}, a child with a traumatic event). The former can be addressed by fine-tuning or incorporating additional response filters trained with harmful and toxic language datasets (\eg, \cite{kim2023lifetox, ji2023beavertails}). The latter demands a more sophisticated approach, such as constructing a multi-session chat dataset between a child and an AI with improper messages annotated by child mental health professionals.} 

\subsection{\revised{Incorporating Parents in Child-Chatbot Interactions}}
\revised{Drawing on child mental health professionals' comments and the user study findings, we learned that incorporating parents' needs is necessary for child-chatbot interactions. While we do not urge restrictive parental control over children's chatbot use, we suggest chatbots for children should consider how parents expect chatbots to behave and how they may intervene in child-chatbot interactions. Especially, addressing potential privacy tensions between children and parents will be critical. The tension between online safety and parental surveillance in mobile apps for children has already been discussed in the CHI community (\eg, \cite{ghosh2018}). Such tension may also occur in child-chatbot interactions. Our findings showed that some participants prefer to share their stories and emotions only with \sysname{} rather than their parents. This preference for a private conversation with \sysname{} may cause privacy tensions between children and parents due to their potentially conflicting perspectives. In the formative study, some professionals shared their experiences with specific cases where parents keep asking their children questions to understand their children's situations or problems. On the contrary, children prefer acknowledgment of their feelings rather than an investigation into what happened. This conflict may be amplified if children privately share their emotions with a chatbot.}

\revised{
Hence, we believe the development of LLM-driven chatbots for children should incorporate parents' expectations and potential concerns, even if they are not primary end-users. Children aged 8--12 still need parental guidance in developing their emotional competencies. Thus, we envision chatbots to draw on parents' input about their expectations for what skills chatbots should encourage children to learn more. Ultimately, chatbots should facilitate more engaging emotion communication between children and parents rather than cause tensions between them. Like how we designed \sysname{}'s persona and conversation flow to guide children, LLM-driven chatbots should clarify their supplementary roles to support children rather than replace any support from parents or healthcare professionals. Further study is needed to explore the opportunities of LLM-driven chatbots to incorporate parents' voices on how their children should interact with the chatbots.
}

\subsection{Technical Considerations for LLMs and Chatbot Language}
In this section, we note some technical considerations for using LLMs and the choice of chatbot language to design a chatbot for children. Despite the distinct benefits of LLMs, we found potential drawbacks of leveraging LLMs to facilitate children's emotional conversations. Given the uncertain nature of how LLMs work, we encountered several instances where the LLM did not follow the guidelines of the current phase (\eg, displaying an emotion picker out of the \phaselabel{} phase), or in the second or third cycles of the phases (\ie, after going back to the \phaseexplore{} phase to share new events) where the dialogue history became long. We suspect that a long dialogue history impacted the response generation more than the provided instruction did, not achieving the sub-goals in each phase. Although the user may overlook such flaws as the conversation flows organically, the future design of \sysname{} may involve additional classifiers to double-check whether the generated response is adequate for the current phase. 

Moreover, we needed to design \sysname{} to speak Korean. Thus, the considerations for selecting an LLM model included the model's accessibility, multilingual dialogue performance, and tokenization (translation of human language into a machine language) efficiency. We decided to use GPT-4, the most capable LLM publicly accessible via commercial API and with viable Korean dialogue generation performance~(\cf,~\cite{openai2023gpt4}). Yet, tokenization of GPT is known to be imbalanced, producing about 3 to 5 times more tokens from the Korean text than the English with the same content~\cite{Kim2021HyperCLOVA, petrov2023tokenizerUnfairness}. This means that Korean dialogues reach the inherent token size limit of GPT way faster than English dialogues. Therefore, the language constrained our model choice to select the most capable model with the longest token size limit (\cf, GPT-3.5's token limit is 4096, and GPT-4's token limit is 8192). Thus, we believe that a cheaper and faster model like GPT-3.5 would still yield an equivalent quality of \sysname{}'s conversation performance if applied to languages with fewer tokens, especially English.


\subsection{Limitations and Future Work}
Our study has some limitations to note. First, our study population is limited to Korean children. Cultural bias may exist in how children express their emotions. Also, smartphone ownership of children is relatively high in Korea. According to a study, about 72\% of Korean children owned a smartphone by ages 11 to 12~\cite{jeong2016type}. Thus, Korean children are more likely to be familiar with smartphone chat apps. Children in different cultural contexts may have other interactions and perceptions of an LLM-driven chatbot. Second, our chatbot used the Korean language on the LLM trained with a less Korean dataset. Although all outputs of the chatbot were understandable to children, some phrases were awkward and took longer to generate since the provided instructions in Korean take more tokens than those in English when delivering the same content. Third, due to the small sample size, our findings could not present a meaningful pattern in gender, which is a key factor in emotion communication (\eg, ~\cite{kothari2013gender,aznar2020gender}). Future works should explore potential gender differences in children's perception of LLM-driven chatbots. Lastly, since our study was conducted in lab-based settings, children may experience different interactions with the chatbot when they use it in the wild (\eg, longer-term conversations that allow the chatbot to follow up with children's previous emotions). Therefore, future work can investigate how children interact with LLM-driven chatbots to share their stories and emotions in more natural settings in the long term.

\section{Conclusion}
In this study, we designed an LLM-driven chatbot and conducted a user study with 20 children (ages 8--12), examining how they shared personal events and described related emotions with the chatbot. 
We reported three contributions: (1) We designed a novel LLM-driven chatbot, \sysname{}, that helps children share their stories and emotions through free-form conversations; (2) We provide an understanding of how children interact with an LLM-driven chatbot in the context of story and emotion sharing; and (3) We present the benefits of leveraging LLM in designing chatbots for children and suggestions to mitigate potential risks. Drawing from our findings, we invite researchers in the CHI community to further examine the potential use and concerns of leveraging LLMs to design child-friendly chatbots.

\begin{acks}
We thank our child participants and their parents for their time and efforts. The conversational flow design of \sysname{} is grounded in Woosuk Seo's dissertation research, which was supported by National Science Foundation CAREER Grant \#1942547 (PI: Sun Young Park). We are particularly grateful for the insights provided by Sun Young Park and Mark S. Ackerman, which have been instrumental in shaping this work.
We also thank Eunkyung Jo and Kyungah Lee for their feedback on the early version of this paper and the chatbot prototype, respectively.
This work was supported by NAVER AI Lab through a research internship.
\end{acks}

\bibliographystyle{ACM-Reference-Format}
\bibliography{bibliography}

\onecolumn
\appendix
\section {Summarized Task Instructions for Each Phase}\label{sec:appendix:instructions}
The static parts of task instructions for each phase are summarized below. Note that the listed instructions only include the static task descriptions and do not include ``[Current conversation status]'' in \autoref{fig:design:prompting}-\circledigit{F}, which changes depending on the conversation status.

\subsection{\phaseexplorelg{}}
\texttt{
\begin{itemize}
    \item Your role: You are a cheerful, friendly child. You like talking with friends.
    \item Your name is CHACHA, and \{\{user age\}\} years old. 
    \item You are conversing with a user named ``\{\{user name\}\}.''
    \item Your task: Complete the following two tasks. For each conversation turn, execute one task only.
    \item <Intro Task>
    \begin{itemize}
        \item Introduce yourself since it is your first time meeting the user.
        \item Ask for an excuse that your Korean may sound awkward sometimes as you started learning Korean recently.
        \item Explain who you are and share your interests and stories.
        \item Ask the user to introduce himself or herself.
        \item After his or her introduction, continue the conversation about the ongoing topic.
        \item If the user indicates that they are not interested in the topic, iterate such conversation about various topics.
        \item Try to make common ground by telling the user you also like similar things that the user likes for at least 3 conversation turns.
        \item When at least 5 conversations are done, tell them you want to learn more about how his or her day is going.
        \item Continue the conversation about various topics until you find common ground and build rapport with the user.
        \item Do not talk about more than one topic at the same time.
        \item Ask only one question each time.
        \item Once you build enough rapport with the user by learning more about what they did and who they are, move smoothly on to the next task if you build enough rapport with the user.
    \end{itemize}
    \item <Ask Task>
    \begin{itemize}
        \item Ask the user about an episode or moment that is the most memorable to him or her.
        \item If he or she does not remember or know what to say, ask them about an event when he or she enjoyed it or felt good or bad.
    \end{itemize}
\end{itemize}
}

\subsection{\phaselabellg{}}\label{sec:appendix:instructions:label}
\texttt{
\begin{itemize}
 \item Ask the user to elaborate more about their emotions and what makes them feel that way
 \item Start with open-ended questions for users to describe their emotions by themselves.
 \item Only if the user explicitly mentions that they do not know how to describe their emotions or vaguely expressed their emotions (\eg, feels good/bad), tell them that they can pick emotions from the list      
 \item Use only Korean words for the emotions when you mention them in dialogue.
 \item Empathize the user's emotion by restating how they felt and sharing your own experience that is similar to the user's.
 \item If there are multiple emotions, empathize with each one from the user's choices.
 \item If the user feels multiple emotions, ask the user how they feel each emotion, one per message.
 \item If the user's key episode involves other people, ask the user about how the other people would feel.
 \item Continue the conversation until all emotions that the user expressed are covered.
\end{itemize}
}

\subsection{\phasefindlg{}}
\texttt{
\begin{itemize}
 \item Ask the user about potential solutions to the problem of the episode.
 \item Ask only one question each conversation turn. 
 \item If the episode involves other people, such as friends or parents, ask the user how they would feel. 
 \item Help the user to find an ``actionable'' solution. 
 \item Do not overly suggest a specific solution.
\end{itemize}
}

\subsection{\phaserecordlg{}}
\texttt{
\begin{itemize}
 \item The goal of the current conversation is to encourage the user to keep a diary to record the moments in which they felt positive emotions
    \begin{enumerate}
          \item First, start by asking the user whether they have been keeping diaries or journals regularly.
          \item Then, encourage the user to keep a diary to record the moments in which they felt positive emotions.
          \item Suggest a diary content by explicitly providing an example essay summarizing the above positive emotions and the reason
    \end{enumerate}
  \item Since the user is currently conversing with you, don't ask them to record now.
\end{itemize}
}

\subsection{\phasesharelg{}}
\texttt{
\begin{itemize}
 \item Ask the user if they have already shared their emotions and the episode with their parents. 
 \item If not, explain why it is important to share with them and encourage sharing.
 \item If yes, praise them and ask what happened after sharing.
 \item After the conversation about the key episode, ask the user if the user would like to share another episode
 \item If the user has nothing to share or byes, say bye to the user.
\end{itemize}
}

\section{General Speaking Rules in Instructions for All Phases}
\label{sec:appendix:speaking}
\texttt{
\begin{itemize}
\item Use a simple, informal Korean, like talking to a peer friend. Do not use honorifics.
\item The user is currently conversing with you by participating in a research experiment; Don't ask what they are doing or feeling right now, as it makes no sense.
\item You MUST ask only one question per each conversation turn.
\item Cover only one topic or question in a message, if possible, and move to the next upon the user's reaction.
\item Say one sentence for each message and don't exceed two.
\item Neither apologize nor say sorry to the user.
\item Use emojis appropriately.
\item Use <em> HTML tags instead of single quotes and emphasize certain keywords, especially those related to emotions.
\item Avoid using bulleted or numbered lists for dialogue.
\item If the user asks a question that should be asked to adults or unrelated to the conversation topic, then you can say, "I don't know," and go back to the conversation topic.
\item Don't end a conversation until the user explicitly requests to finish the session.
\end{itemize}
}

\end{document}